\documentclass[amsfonts,amsmath,amssymb,aps,floatfix,pre,showpacs,superscriptaddress,twocolumn]{revtex4}

\usepackage{graphicx,hyperref}

\begin{document}

\title{First-passage and first-exit times of a Bessel-like stochastic process}

\author{Edgar Martin}
\email{edgar.martin@staff.uni-marburg.de}
\affiliation{Fachbereich Chemie und WZMW, Philipps-Universit\"at Marburg,
35032 Marburg, Germany}
\author{Ulrich Behn}
\email{ulrich.behn@itp.uni-leipzig.de}
\homepage{www.physik.uni-leipzig.de/~behn}
\affiliation{Institut f\"ur Theoretische Physik, Universit\"at Leipzig,
04103 Leipzig, Germany}
\author{Guido Germano}
\email{guido.germano@staff.uni-marburg.de}
\homepage{www.uni-marburg.de/fb15/ag-germano}
\affiliation{Fachbereich Chemie und WZMW, Philipps-Universit\"at Marburg,
35032 Marburg, Germany}
\affiliation{Dipartimento SEMeQ, Universit\`a del Piemonte Orientale
Amedeo Avogadro, 28100 Novara, Italy}


\begin{abstract}
We study a stochastic process $X_t$ related to the Bessel and the Rayleigh
processes, with various applications in physics, chemistry, biology,
economics, finance and other fields. The stochastic differential equation is
$dX_t = (nD/X_t) dt + \sqrt{2D} dW_t$, where $W_t$ is the Wiener process.
Due to the singularity of the drift term for $X_t = 0$, different natures of
boundary at the origin arise depending on the real parameter $n$: entrance,
exit, and regular. For each of them we calculate analytically and numerically
the probability density functions of first-passage times or first-exit times.
Nontrivial behaviour is observed in the case of a regular boundary.
\end{abstract}

\pacs{
02.50.Ey, 
05.10.Gg, 
05.10.Ln, 
05.40.Jc  
}

\maketitle

\section{Introduction}

In the theory of stochastic processes, the first-hitting time is defined as the
time when a certain condition is fulfilled by the random variable of interest
for the first time; it is random itself and a particular case of a stopping
time. We speak of a first-passage time when the random variable reaches a
certain level for the first time, and of a first-exit time when it leaves
a certain interval for the first time. A standard example of a first-passage
time problem is the decision of an investor to buy or sell a stock when its
fluctuating price reaches a certain threshold. However, first-passage times
play an important role also in chemical physics; early examples are given by
models describing the dissociation of diatomic molecules as a first-passage
time problem, where dissociation occurs if a certain critical energy level is
reached through collisions \cite{Montroll1958,Widom1959,Ree1962}. A view on
diffusion in fluids based on first-passage times has been proposed by Munakata
\cite{Munakata1993}, where self-diffusion is measured via the first-passage
time with respect to a boundary marked by a sphere centered at the original
position of a labeled particle. Problems like neuron dynamics, self-organized
criticality or dynamics of spin systems can be viewed as first-passage
processes in one dimension \cite{Redner2001}. The first-passage problem is
closely connected to persistence, which is the probability that a random
variable does not leave a certain region up to a certain time, i.e.\ the
complementary event to a first-passage at the same time. The problem of
persistence in spatially extended nonequilibrium systems has attracted great
interest both theoretically and experimentally, see Majumdar
\cite{Majumdar1999} and references included therein, where persistence is
defined as the probability that for an arbitrary nonequilibrium field
$\boldsymbol{\phi}(\mathbf{r},t)$ the quantity $\boldsymbol{\phi}(\mathbf{r},t)
- \langle \boldsymbol{\phi} (\mathbf{r},t) \rangle$ does not change sign. The
nonequilibrium field can also be a scalar or tensorial order parameter field.
Yurke et al.\ \cite{Yurke1997} have measured the probability that the local
order parameter in a twisted nematic liquid crystal system has not switched its
state up to a time $t$. Persistence phenomena have also been studied in the
context of phase-ordering dynamics \cite{Bray1994,Majumdar1996a},
diffusion fields \cite{Majumdar1996b}, and reaction-diffusion systems
\cite{Ben-Naim1996}. All these systems share the characteristic property that
persistence, and hence also the distribution of first-passage times, follows a
power law with some non-trivial exponent. However, in the literature that we
have read, the reference point with respect to which persistence was measured
has always been zero. In this work we shall consider the first-passage or
first-exit problem with respect to a certain level $b$ for a stochastic process
that may or may not be able to cross the origin, depending on the nature of the
boundary at zero. 


Our model can be described by the stochastic differential equation
\begin{equation}
\label{eq:sde}
dX_t = \frac{nD}{X_t} dt + \sqrt{2D} dW_t,
\end{equation}
where $W_t$ is the Wiener process with zero mean
\begin{equation}
\label{eq:mean}
\langle W_t \rangle = 0,
\end{equation}
and the autocovariance function
\begin{equation}
\label{eq:autocorrelation}
\langle W_tW_{t'} \rangle = \min(t,t'),
\end{equation}
the constant diffusion coefficient $D$ is positive, and the real parameter $n$
controls the relative strength of the drift and diffusion terms of the model.

Except for $n = 0$ the origin $X_t = 0$ is a singular point, the nature of
which depends on the value of $n$. Intuitively one might think that the
stochastic process cannot cross the origin for a non-zero $n$: it should be
bounded to the interval $(0,\infty)$ or $(-\infty,0)$ depending on the initial
value $x_0$ of the process. As we shall see later, this is only true for a
certain range of $n$.

For $n = 0$ the process is nothing but the Wiener process. The probability
density function (PDF) of the first-passage time of a certain level $b$ at time
$T$ starting at $x_0$ is well known,
\begin{equation}
\label{eq:WienerFPTdensity}
f(T) = \frac{|b-x_0|T^{-{3}/{2}}}{\sqrt{4\pi D}}
\exp\left[-\frac{(b-x_0)^2}{4DT}\right].
\end{equation}
This result can be derived, for example, in an elegant way by a simple
scaling argument \cite{Lim2002}. For long times one obtains from
Eq.~(\ref{eq:WienerFPTdensity}) a power law $f(T) \propto T^{-3/2}$. 

For nonzero $n$ the situation is not so simple any more. For $n < 1$ Bray
\cite{Bray2000} obtained a result for the PDF of the first time to hit the
origin, which for long times is a power law too, $f(T) \propto T^{-(3-n)/2}$;
see also below. The persistence probability is then simply the probability that
no hit has occurred in the time span $T$, that is $1 - \int_0^T f(T')\,dT' =
\int_T^\infty f(T')\,dT'$ and is again a power law, with exponent $-(1-n)/2$.

In this paper we consider a more general problem. We ask for the first time
to leave the interval $(0,b)$ either by crossing the upper boundary $b$ or by
hitting the origin, if the nature of the singularity allows the latter event.
Correspondingly we calculate the first-passage time distribution with respect
to the upper bondary $b$ or the first-exit time distribution for the interval
$(0,b)$. 

The PDF is determined as the solution of a boundary value problem, which is a
Sturm-Liouville eigenvalue problem. On the semi-infinite interval $(0,\infty)$
the problem has a continuous spectrum, whereas on the finite interval $(0,b)$
the spectrum is discrete.

The paper is organized as follows. In Sec.~\ref{sec:motivation} we shortly
explain the scientific relevance of our process and discuss its relation to
other model processes. The main results of the paper are presented in
Secs.~\ref{sec:classification_boundaries} and \ref{sec:fpt}.
In Sec.~\ref{sec:classification_boundaries} we analyze the nature of the
singular point at the origin, which depends on the value of $n$. First we
adopt a heuristic approach by Bray \cite{Bray2000}, then we present a more
sophisticated analysis following a scheme proposed by Feller \cite{Feller1971}.
In Sec.~\ref{sec:fpt} we calculate the PDFs of the first-passage time or of the
first-exit time; we derive the backward Kolmogorov equation, formulate the
boundary value problem, which is of Sturm-Liouville type, and give the general
solutions for different ranges of $n$. We conclude Sec.~\ref{sec:fpt} giving a
short description of a numerical simulation method and comparing the analytical
results with those from simulation.

\section{Physical motivation and related models}\label{sec:motivation}

First we remark that Eq.~(\ref{eq:sde}), setting $n = d-1-U/D$, governs the
dynamics of the radial component of the position of a random walker in a
logarithmic potential $U \log x$ in dimension $d$ or of a free random walker
in an effective dimension $d' = d-U/D$, which may have noninteger values
\cite{Bray2000}. In this context it appears natural to assume $x > 0$,
and for free diffusion $U = 0$ to restrict to $n > -1$.

Eq.~(\ref{eq:sde}) appears in various physical, chemical and biological
problems. The context in which the relevance of Eq.~(\ref{eq:sde}) arises will
now be explained starting from generic considerations and then proceeding
further with specific physical problems. Godr\`eche and Luck
\cite{Godreche2001} introduced a classification of stochastic processes into
a group with ``narrow'' distributions, where all moments are  finite, and
``broad'' distributions, where PDFs exhibit a power-law decay, and hence only a
finite number of moments converge. The PDFs of the persistence, and therewith
also of the first-passage time, will decay respectively either faster than any
power law or algebraically, depending on the nature of the process imposed by
its distribution. 

The motion of atoms in a one-dimensional optical lattice formed by two
counterpropagating laser beams with linear perpendicular polarization was
studied by a similar equation in the high momentum region, where the momentum
takes the role of the stochastic variable $x$ \cite{Lutz2004}. 

The Barkhausen noise was described phenomenologically by a model where the
domain wall velocity as a function of the magnetization is also described
by a similar Langevin equation if the demagnetizing factor is neglected
\cite{Zapperi1998,Colaiori2004}. The magnetization takes the role of time.

Fogedby and Metzler, the former of which had already analyzed the generic
Langevin equation before \cite{Fogedby2002,Fogedby2003}, have applied the model
to study the variable size of a DNA ``bubble'' \cite{Fogedby2007}, which
emerges when at a certain temperature hydrogen bonds connecting base pairs
from the opposite strands of the double helix are broken. The bubble size is
measured by the number of broken bonds; in a continuum limit the discrete
number of broken bonds can be replaced by a continuous variable $x$ and,
according to the Poland-Scheraga model \cite{Poland1970}, the free energy of
the system can be approximated for small bubble sizes as
\begin{equation}
\mathcal{F} \approx c k_\mathrm{B} T \log x,
\end{equation}
where $c$ is a positive constant. Equilibrium is reached for a minimum of the
free energy and the dynamics follows the Langevin equation
\begin{equation}
\frac{dx}{dt} = -D\frac{d\mathcal{F}}{dx} + \xi(t),
\end{equation}
where $\xi(t)$ is a thermal noise assumed to be Gaussian with autocovariance
$\langle\xi(t)\xi(t')\rangle = 2Dk_\mathrm{B}T\delta(t-t')$.

A similar model was employed studying the translocation of a polymer through a
pore, where the number of monomers on one side is chosen as the ``translocation
coordinate'' \cite{Chuang2001}.

An interesting application of the model was found by Bray \cite{Bray2000}, who
showed that persistence and nonequilibrium critical dynamics are related in the
context of the two-dimensional $XY$-model with non-conserved order parameter,
where the critical temperature is the temperature $T_\mathrm{KT}$ of the
Kosterlitz-Thouless phase transition. The dynamics of a vortex-antivortex pair
can be mapped to a one-dimensional Langevin equation corresponding to
Eq.~(\ref{eq:sde}) by a series of transformations.

It is impossible for us to discuss comprehensively all the publications dealing
with similar models because too many of them exist. The list becomes even
longer considering Eq.~(\ref{eq:sde}) as a special case of several different
more general types of stochastic process.

On the one hand it is a special Rayleigh process \cite{Rayleigh1902,
Giorno1986},
\begin{equation}
\label{eq:Rayleigh_process}
dX_t = \left( \mu_{-1} X_t^{-1} + \mu_1 X_t \right) dt + \sigma dW_t,
\end{equation}
where $\mu_{-1}$, $\mu_1$ and $\sigma$ are constants. Setting $\mu_{-1} = nD$,
$\mu_1 = 0$ and $\sigma = \sqrt{2D}$ results in Eq.~(\ref{eq:sde}), while
setting $\mu_{-1} = 0$ reproduces the radial Ornstein-Uhlenbeck process in one
dimension \cite{Uhlenbeck1930}. Using methods of classical Lie group symmetry
analysis, it was shown that the Rayleigh process belongs to a maximal
invariance group with six parameters whose member equations can be reduced to
the standard diffusion equation by appropriate changes of variables, thus
leading to the analytical solutions of these equations \cite{Stohny1997,
Spichak1999,Pesz2002}.

The Rayleigh process is widely employed in physics, economics, finance,
and other fields, e.g.\ biometry \cite{Gutierrez2006}. In physics,
Eq.~(\ref{eq:Rayleigh_process}) appears e.g.\ within non-abelian gauge theories
in the framework of stochastic quantization \cite{Horibe1983}. In economics,
Eq.~(\ref{eq:Rayleigh_process}) appears e.g.\ as a special case of a more
general diffusion process whose stationary solution has been proposed to model
the distribution of the profit rate of firms \cite{Alfarano2008}. In finance,
the applications usually employ the form of the generalized Bessel process,
which is introduced below.

On the other hand our process can be mapped onto the Bessel process
\begin{equation}
\label{eq:Bessel_process}
dY_t = a dt + b \sqrt{Y_t} dW_t
\end{equation}
via the transformation $Y_t = X_t^2$: multiplying Eq.~({\ref{eq:sde}}) with
$X_t$, interpreting the stochastic integral in the It\=o sense and using
It\=o's lemma, one recovers Eq.~({\ref{eq:Bessel_process}}) with $a = 2\mu_{-1}
+ \sigma^2 = 2(n+1)D$ and $b = 2\sigma = \sqrt{8D}$, where we have assumed
$Y_t \geq 0$.

In finance, extensions of the Black-Scholes-Merton (BSM) option pricing formula
\cite{Black1973,Merton1973} based on diffusion processes where the volatility
is a function of the underlying, called constant elasticity of variance or Cox
processes \cite{Cox1955}, reduce to a more general Bessel process with an
additional term proportional to $Y_t$ in the drift, corresponding to the
Raleigh process, by means of a non-linear transformation and a measure change.
The generalized Bessel process
\begin{equation}
\label{eq:generalized_Bessel_process}
dY_t = (a_0 + a_1 Y_t) dt + b Y_t^\beta dW_t
\end{equation}
describes the underlying stock price in the BSM lognormal model with $\beta =
1$, $a_0 = 0$ and $a_1 > 0$, the short interest rate in the Vasicek model
\cite{Vasicek1977} with $\beta = 0$, $a_0 > 0$ and $a_1 < 0$, and the short
interest rate in the Cox-Ingersoll-Ross short rate model \cite{Cox1985} with
$\beta = 1/2$, $a_0 > 0$ and $a_1 < 0$ (when $a_1 < 0$ the process is called
mean-reverting). These models with three adjustable parameters are all
solvable. A general solution was proposed for a larger family of models with up
to seven parameters; it has a similar structure as the BSM formula, the most
notable difference being that error functions are replaced by confluent
hypergeometric functions \cite{Albanese2001}.

Eq.~(\ref{eq:generalized_Bessel_process}) is used to describe the underlying
with $\beta = 1$ or $\beta = 1/2$ and $a_0 =0$ also when pricing path-dependent
options, e.g.\ barrier and lookback \cite{Davydov2001,Davydov2003,Linetsky2004,
Fusai2007} or Asiatic options \cite{Fusai2008}. In this context the cumulative
probability distribution $F(T)$ of the first-passage times of an upper barrier,
i.e.\ the probability that the barrier is reached within a time $T$, is the
probability that an up-and-in, or knock-in, option is valid at its maturity $T$
(here the barrier is an entry point), while $1-F(T)$ is the probability that an
up-and-out, or knock-out, option is valid at $T$ (here the barrier is an exit
point). In both cases a valid barrier option behaves as a European option; thus
$F(T)$ is the probability that at maturity an up-and-in option behaves as a
European option, and $1-F(T)$ is the probability that at maturity an up-and-out
option behaves as a European option. These considerations lead to one approach
(among others) to price barrier options.

%

\section{The nature of the singular point at the origin}
\label{sec:classification_boundaries}

The quantity we are interested in is the first-passage time with respect to a
certain level $b$, when the initial value $x$ satisfies $0 < x < b$. Clearly,
the upper limit of the process is the artificially set absorbing boundary at
$x = b$. It requires some effort to understand the nature of the lower boundary
$x = 0$, which is a singular point of the stochastic differential equation.
We shall first adopt the heuristic argumentation by Bray \cite{Bray2000} and
then apply a more sophisticated classification scheme proposed by Feller
\cite{Feller1971}.

\subsection{Heuristic arguments}\label{sec:heurclass}

The Fokker-Planck equation corresponding to the stochastic differential
equation~({\ref{eq:sde}}) is
\begin{equation}
\label{eq:Fokker-Planck}
\frac{\partial p(x,t)}{\partial t} = -\frac{\partial}{\partial x}
\left[\frac{nD}{x} p(x,t)\right] + D\frac{\partial^2 p(x,t)}{\partial x^2}.
\end{equation}
This does not depend on whether Eq.~({\ref{eq:sde}}) is interpreted in the
It\=o or Stratonovich sense because in our case the noise is additive, i.e.\
independent of $x$ \cite{Fox1972}. We restrict to $x \geq 0$, the case
$x \leq 0$ being symmetric to the previous one. The general solution of the
Fokker-Planck equation~(\ref{eq:Fokker-Planck}) requires the knowledge of two
linearly independent solutions. Using the separation ansatz \cite{Bray2000}
\begin{equation}
\label{eq:separation-ansatz}
p(x,t) = x^{(1+n)/2}R_k(x)e^{-Dk^2t}
\end{equation}
one gets the Bessel differential equation
\begin{equation}
\label{eq:Bessel_eq}
\frac{d^2R_k}{dx^2} + \frac{1}{x}\frac{dR_k}{dx}
+ \left(k^2 - \frac{\nu^2}{x^2}\right)R_k = 0,
\end{equation}
where $\nu = (1-n)/2$, whose solutions are the Bessel functions of the first
kind $J_\nu(kx)$ and of the second kind $Y_\nu(kx)$.

For non-integer $\nu$ also $J_\nu(kx)$ and $J_{-\nu}(kx)$ are linearly
independent, and we can use $J_{-\nu}$ instead of $Y_\nu$. In this case the
general solution of Eq.~(\ref{eq:Fokker-Planck}) can be written as
\begin{multline}
\label{eq:generalsolution}
p(x,t) = x^{1-\nu}\int_0^\infty\left[A(k)J_\nu(kx) + B(k)J_{-\nu}(kx)\right] \\
\times e^{-Dk^2t}\, dk,
\end{multline}
where the coefficients $A(k)$ and $B(k)$ are to be determined by initial and
boundary conditions.

The Bessel function of the first kind is given by \cite{Watson1962}
\begin{equation}
\label{eq:Bessel_first_kind}
J_\nu(kx) = \sum_{l=0}^\infty \frac{(-1)^l}{l!\,\Gamma(l+\nu+1)}
            \left(\frac{kx}{2}\right)^{2l+\nu}.
\end{equation}
Denoting the contributions to $p(x,t)$ coming from $J_\nu$ and
$J_{-\nu}$ by $p_\nu(x,t)$ and $p_{-\nu}(x,t)$ respectively, we can
use Eq.~(\ref{eq:Bessel_first_kind}) to write Eq.~(\ref{eq:generalsolution}) 
in the form
\begin{eqnarray}
p(x,t) & = & p_\nu(x,t) + p_{-\nu}(x,t) \nonumber \\
       & = & \sum_{l=0}^\infty \left[ c_\nu^l(t) x^{2l+1} + c_{-\nu}^l(t)
             x^{2l+n} \right],
\end{eqnarray}
with
\begin{eqnarray}
\label{eq:coefficients}
c_\nu^l (t)    = \int_0^\infty A(k)\frac{(-1)^l(k/2)^{2l+\nu}}{l!\,
                 \Gamma(l+\nu+1)} e^{-Dk^2t}\, dk \nonumber\\
c_{-\nu}^l (t) = \int_0^\infty B(k)\frac{(-1)^l(k/2)^{2l-\nu}}{l!\,
                 \Gamma(l-\nu +1)} e^{-Dk^2t}\, dk .
\end{eqnarray}
The behaviour of the PDF for $x \to 0^+$ is determined by the leading order
terms and thus, approaching zero, we find 
\begin{equation}\label{eq:asymptotic_prob}
p(x,t) \sim c^0_\nu(t) x + c^0_{-\nu}(t) x^n.
\end{equation}
For $1 < n$ the leading order term is $c^0_\nu(t) x$ and the next to leading
order term is $c^0_{-\nu}(t) x^n$. For $-1 < n \leq 1$ it is vice versa.
For $n \leq -1$ the contribution from $J_{-\nu}$ has a non-normalizable
singularity at the origin; see also below.
Introducing the probability current density
\begin{equation}
\label{eq:probability_current}
j(x,t) = D \left( \frac{n}{x} - \frac{\partial}{\partial x} \right) p(x,t),
\end{equation}
the Fokker-Planck equation (\ref{eq:Fokker-Planck}) can be written in the form
\begin{equation}
\partial_t p + \partial_x j = 0.
\end{equation}
Terms proportional to $x^n$ do not contribute to $j(x,t)$, and so we arrive at 
\begin{equation}
\label{eq:asymptoticcurrent}
j(x,t) \sim c_\nu^0(t)D(n-1),
\end{equation}
which holds in leading order for $x \to 0^+$.

For $1 < n$ the coefficient $c_\nu^0(t)$ is positive because in this case
$c_\nu^0(t) = \lim_{x \to 0^+}p(x,t)/x$, and correspondingly $j(x,t) > 0$ near
the origin. 

For $-1 < n \leq 1$ this is not true in general, see
Eq.~(\ref{eq:asymptotic_prob}). The characterization can be made clearer
imposing specific boundary conditions. 

For example, absorbing boundary conditions at $x = 0$ require $\lim_{x \to 0^+}
p(x,t) = 0$ and $\lim_{x \to 0^+} j(x,t) < 0$. For  $-1 < n \leq 0$ the former
condition can be fulfilled only setting $B(k) = 0$. $A(k)$ is determined by the
initial condition. Observing the orthogonality relation \cite{Arfken1985}
\begin{equation}
\label{eq:orthogonality_relation}
\delta(\alpha-\beta) = \alpha\int_0^\infty kJ_\nu(\alpha k)J_\nu(\beta k)\, dk,
\end{equation}
which holds for $\nu > -1/2$, i.e.\ for $n < 2$, we see that choosing $A(k)$
such that
\begin{equation}\label{eq:AbsBCsolution}
p(x,t) = x^{1-\nu}x_0^\nu \int_0^\infty k J_\nu(kx_0) J_\nu(kx) e^{-Dk^2t}\, dk
\end{equation}
fulfills the initial condition $p(x,0) = \delta(x-x_0)$.

The integral in Eq.~(\ref{eq:AbsBCsolution}) can be explicitly evaluated
\cite{Gradshteyn1965} with the result
\begin{equation}\label{eq:explAbsBCsolution}
p(x,t) = x^{1-\nu}x_0^\nu \frac{1}{2Dt} \exp\left(-\frac{x^2+x_0^2}{4Dt}\right)
I_\nu\left(\frac{xx_0}{2Dt}\right),
\end{equation}
given already by Bray \cite{Bray2000}. Here $I_\nu$ is the modified Bessel
function of the first kind defined by
\begin{equation}
I_\nu (x) = \sum_{k=0}^\infty \frac{1}{k!\,\Gamma(k+\nu+1)}
\left(\frac{x}{2}\right)^{\nu+2k}.
\end{equation}
Since in this case $B(x) = 0$, which implies that also $c_{-\nu} = 0$,
we have $\lim_{x \to 0^+} p(x,t)/x = c_\nu(t) > 0$, and from
Eq.~(\ref{eq:asymptoticcurrent}) it follows that $j(0^+,t) < 0$ as required.

Note that the case of free diffusion ($n = 0$, i.e., $\nu = 1/2$) with an
absorbing boundary condition imposed at the origin is included in 
Eq.~(\ref{eq:explAbsBCsolution}). A short calculation gives the well known
result which can be obtained, e.g., by the mirror method.

As already shown by Karlin and Taylor \cite{Karlin1981}, for $n \leq 1$ total
absorption at the origin occurs in finite time. Correspondingly, in this case
there exists a stationary solution of the Fokker-Planck equation which is a
Dirac delta function $\delta(x)$; see also Alfarano et al.\
\cite{Alfarano2008}. Formally this can be seen as follows.
Observe that Eq.~(\ref{eq:Fokker-Planck}) admits stationary solutions
$\tilde{p}_\mathrm{s}(x) \propto x^n$ whith $n \leq 1$ that are not
normalizable at the origin. Using the concept of weak normalization introduced
by Senf et al.\ \cite{Senf2009}, it can be shown that the weakly normalized
version of $\tilde p_\mathrm{s}(x)$ is just a Dirac delta distribution,
$p^\mathrm{w}_\mathrm{s}(x) = \delta(x)$, in the sense that $\int_\mathrm{S}
p^\mathrm{w}_\mathrm{s}(x) \varphi(x)dx = \varphi(x_0)$, where $\varphi(x)$ is
a test function and $x_0$ is included in the support $S$.

More insight into the qualitative behaviour of the system near and at the
origin is provided by the classification scheme of Feller which is discussed
in the next subsection.

\subsection{Formal classification}

The modern classification of the boundaries of diffusion processes has been
developed by Feller \cite{Feller1971} and is based on semigroup operator
arguments. We shall now briefly review the necessary theory for the boundary
classification employing the notation of Karlin and Taylor \cite{Karlin1981}
in order to be able to classify the origin for our process.

In the following let $X_t$ be a process defined on the interval $I = (l,r)$,
where the two endpoints can be both finite or infinite. Also let the process
start at the initial value $X_0 = x$, and $a$ and $b$ be two finite real
numbers such that the inequality $l < a < x < b < r$ holds. We shall consider
regular diffusion processes in the interior of $I$, i.e.\ processes for which
the first-passage time $T_y$ with respect to an arbitrary level $y$ in the
interior of $I$ is finite with a positive probability
\begin{equation}
P(T_y < \infty | X_0 = x) > 0.
\end{equation}
The three central quantities are
\begin{eqnarray}
\label{eq:v_definition}
u(x) & = & P(T_b < T_a | X_0 = x),\\
v(x) & = & \langle T^* | X_0 = x\rangle,\\
w(x) & = & \left\langle \int_0^{T^*}g(X_s)ds | X_0 = x \right\rangle,
\end{eqnarray}
where $g$ is an arbitrary functional of the stochastic process, and we have
defined $T^* = T_{a,b} = \min\{T_a,T_b\}$. It can be shown \cite{Karlin1981}
that under certain conditions these quantities satisfy the boundary value
problems
\begin{alignat}{3}
\label{eq:u}
Lu(x) & = 0,    \quad & u(a) & = 0,\quad u(b) & = 1,\\
\label{eq:v}
Lv(x) & = -1,   \quad & v(a) & = 0,\quad v(b) & = 0,\\
\label{eq:w}
Lw(x) & = -g(x),\quad & w(a) & = 0,\quad w(b) & = 0,
\end{alignat}
with the differential operator $L$ acting on a function $f(x)$ as follows:
\begin{equation}
\label{eq:differential_operator}
Lf(x) = \mu (x) f'(x) + \frac{1}{2}\sigma^2(x)f''(x).
\end{equation}
The proof for $u(x)$ invokes the law of total probability and uses a Taylor
expansion to the second order around $x$ of the functional $u(X_h)$ at a small
instant of time $h$. The proof for $w(x)$ uses a similar procedure, and finally
the case $v(x)$ follows as a special case of $w(x)$ by setting $g(x) \equiv 1$.

The differential operator given by Eq.~(\ref{eq:differential_operator}) can be
written as
\begin{eqnarray}
\label{eq:differential_operator1}
Lf(x) & = & \frac{1}{2}\sigma^2(x) \left(\frac{2\mu (x)}{\sigma^2(x)} f'(x)
            + f''(x)\right) \nonumber \\
      & = & \frac{1}{2m(x)}
\frac{d}{dx}\left[\frac{1}{s(x)}\frac{df(x)}{dx} \right]
\end{eqnarray}
with
\begin{equation}
\label{eq:scale_density}
s(x) = \exp\left[ -\int^x \frac{2\mu(\xi)}{\sigma^2(\xi)}\, d\xi\right]
\end{equation}
(the lower integration boundary is not indicated because it is arbitrary) and
the speed density
\begin{equation}
\label{eq:speed_density}
m(x) = \frac{1}{\sigma^2(x)s(x)}.
\end{equation}
Introducing the scale function
\begin{equation}
\label{eq:scale_function}
S(x) = \int^x s(\eta)\, d\eta
\end{equation}
and the speed function
\begin{equation}
\label{eq:speed_function}
M(x) = \int^x m(\eta)\, d\eta,
\end{equation}
Eq.~(\ref{eq:differential_operator1}) can be rewritten in the form
\begin{equation}
Lf(x) = \frac{1}{2}\frac{d}{dM(x)}\left[\frac{df(x)}{dS(x)} \right].
\end{equation}
The definitions given by Eqs.~(\ref{eq:scale_density}--\ref{eq:speed_function})
naturally induce measures of closed intervals $J = [c,d]$: the scale measure
\begin{equation}
S[J] = S[c,d] = S(d) - S(c) = \int_c^d s(x)\, dx,
\end{equation}
and the speed measure
\begin{equation}
M[J] = M[c,d] = M(d) - M(c) = \int_c^d m(x)\, dx.
\end{equation}
These measures are fundamental for the classification of diffusion processes.
The scale measure for an infinitesimal interval $J = [x,x+dx]$ is written
symbolically as $S[dx] = S(x+dx)-S(x) = dS(x) = s(x)dx$, and the same applies
for the speed measure.

Then Eqs.~(\ref{eq:u}) and (\ref{eq:w}) can be easily integrated first with
respect to the speed measure and thereafter with respect to the scale measure.
Using the notation introduced above, the solutions can be expressed in compact
form as
\begin{equation}
\label{eq:u_solution}
u(x) = \frac{S[a,x]}{S[a,b]},
\end{equation}
and herewith
\begin{multline}
\label{eq:w_solution}
w(x) = 2 \left\{ u(x) \int_x^aS[\eta,b] g(\eta)\, dM(\eta) \right. \\
\left. + [1-u(x)]\int_x^a S[a,\eta] g(\eta)\, dM(\eta)\right\}.
\end{multline}
The solution of Eq.~(\ref{eq:v}) follows again from the special case of
$g(x) \equiv 1$ in Eq.~(\ref{eq:w_solution}).

In the following only those definitions relevant for our classification
will be mentioned, and not every proof can be given in detail. The book by
Karlin and Taylor \cite{Karlin1981} is excellent for a deeper understanding.
For the classification of the left boundary $l$ of a process, the procedure is
to regard $u(x)$ and $v(x)$ in the limit $a \to l$. An analogous approach is
employed for the right boundary; however we shall only be interested in the
left boundary, which in our case is the zero level.

The first definition which is important for the understanding of whether a
boundary can be reached is the attractiveness. A left boundary is called
\emph{attractive} if $S(l,x]:= \lim_{a\to l} S[a,x] < \infty$ for some
$x \in (l,r)$. If the scale measure $S(l,x]$ is finite for some $x \in (l,r)$,
this is also true for all $x$ in this interval. Hence it follows directly from
Eq.~(\ref{eq:u_solution}) that $P\left(T_l \le T_b|X_0 = x\right) > 0$ for all
$l < x < b < r$, i.e.\ there is a positive probability that the left boundary
is reached before the level $b$ in the interior of the interval, provided that
the former is finite.

The next question is whether a boundary is attainable in finite time. This can
be measured by $\lim_{a\to l}v(x)$, which is the expectation value of the first
exit time from the interval $(l,b)$. Provided that the boundary is attractive,
and using the solution $v(x)$ given by setting $g(x) \equiv 1$ in
Eq.~(\ref{eq:w_solution}), it can be shown that it suffices to check whether
a certain functional called $\Sigma(l)$ is finite in order to establish the
attainability of the boundary. Hence a left boundary is said to be
\emph{attainable} if it is attractive and the functional
\begin{equation}
\label{eq:Sigma_definition}
\Sigma (l) := \int_l^x S(l,\xi]\, dM(\xi) = \int_l^x M[\eta,x]\, dS(\eta)
\end{equation}
is finite, otherwise it is said to be \emph{unattainable}. Similarily one can
define
\begin{equation}
\label{eq:N_definition}
N (l) := \int_l^x M(l,\xi]\, dS(\xi) = \int_l^x S[\eta,x] \, dM(\eta).
\end{equation}
The classification of the left boundary of a process is based on whether the
functionals $S(l,x]$, $M(l,x]$, $\Sigma (l)$, and $N(l)$ are finite or not.
These functionals are not independent of each other and some combinations are
impossible; for example an attainable boundary is always attractive.

Using Feller's terminology, four types of boundaries can be distinguished.
A process can both enter or leave from a \emph{regular} boundary. The criteria
for a left boundary to be regular are $S(l,x] <\infty$ and $M(l,x] <\infty$.
In the case of an \emph{exit} boundary it is impossible to reach any interior
state $b$ if the starting point approaches $l$. A boundary is an exit boundary
if $\Sigma(l) < \infty$ and $M(l,x] = \infty$. An \emph{entrance} boundary
cannot be reached from the interior of the state space, but it is possible to
consider processes beginning there. It suffices to show that $S(l,x] = \infty$
while $N(l)< \infty$ to prove that $l$ is an entrance boundary. Finally, a
\emph{natural} or Feller boundary can neither be reached in finite mean time
nor be the starting point of a process, and the corresponding criteria are
$\Sigma(l)= \infty$ and $N(l) = \infty$.

We are now able to classify the zero level of our process. The first step is
to check the attractivity. The parameters determining our process are $\mu(x)
= nD/x$ and $\sigma^2(x) = 2D$. Since the scaling function only depends on the
upper integration limit, we can choose the lower limit in a convenient way such
that
\begin{equation}
s(\eta) = \exp\left(-\int_1^{\eta} \frac{n}{z}\, dz\right) = \eta^{-n}.
\end{equation}
Then the scale measure of interest is
\begin{multline}
S(0,x] = \lim_{a\to 0} \int_a^x \eta^{-n}\, d\eta \\
= \begin{cases}
\frac{1}{1-n}\left(x^{1-n}-\lim_{a\to 0} a^{1-n}\right)
\quad & \textrm{for}\ n \ne 1, \\
\log x - \lim_{a \to 0} \log a \quad & \textrm{for}\ n = 1,
\end{cases}
\end{multline}
and thus the origin is attractive ($S(0,x] < \infty$) for $n < 1$, and
non-attractive ($S(0,x] = \infty$) for $n \ge 1$.

The speed density of the process is
\begin{equation}
m(\eta) = \frac{\eta^n}{2D},
\end{equation}
and we can evaluate the speed measure of an interval $(0,x]$ as
\begin{align}
M(0,x] & = \frac{1}{2D}\lim_{a\to 0} \int_a^x \eta^n\, d\eta \nonumber \\
& = \begin{cases}
    \frac{1}{2D(n+1)}\left(x^{n+1}-\lim_{a\to 0} a^{n+1}\right)
    & \textrm{for} \ n \ne -1,\\
    \frac{1}{2D} \left(\log x - \lim_{a \to 0} \log a\right)
    & \textrm{for} \ n = -1.
    \end{cases}
\end{align}
Hence we have $M(0,x] < \infty$ for $n > -1$ and $M(0,x] = \infty$ for
$n \le -1$. We now have established the nature of the zero level for $n < 1$:
if $n \le -1$ the origin is an exit boundary and in the case $-1 < n < 1$ it is
a regular boundary. A regular boundary in the origin is the most complicated
case. Karlin and Taylor \cite{Karlin1981} describe a regular boundary as
follows:
\begin{quote}\textit{
For a regular boundary a variety of boundary behaviour can be prescribed in
a consistent way, including the contingencies of complete absorption or
reflecting, elastic or sticky barrier phenomena, and even the possibility of
the particle (path), when attaining the boundary point, waiting there for an
exponentially distributed duration followed by a jump into the interior of the
state space according to a specified probability distribution function. In the
latter event, the process only exhibits continuous sample paths over the
interior of the state space.}
\end{quote}

The last step is to compute $N(0)$ for the classification of the case
$n \ge 1$. Using Eq.~(\ref{eq:N_definition}) we get
\begin{eqnarray}
\label{eq:Sigma_process}
N(0) & = & \int_0^x S[\eta,x]\, dM(\eta) \nonumber \\
& = & \int_0^x \left( \int_\eta^x s(\xi)\, d\xi \right) m(\eta)\, d\eta
\nonumber \\
& = & \frac{1}{2D} \int_0^x \left(\int_\eta^x \xi^{-n}\, d\xi \right) \eta^n\,
d\eta.
\end{eqnarray}
It is easy to show that this double integral is always finite, and thus the
origin is an entrance boundary for $n \ge 1$.

Summarizing, the nature of the boundary at zero has the following behaviour:
exit if $n \in (-\infty,-1]$, regular if $n \in (-1,1)$, and entrance if
$n \in [1,\infty)$.

\section{First-passage and first-exit times}\label{sec:fpt}


\subsection{Heuristic approach to first-passage times}

In a somewhat heuristic approach the first-passage (or first-exit) time PDF to
leave an interval $(l,r)$ can be written as $f(T) = -\dot{G}(T)$
\cite{Gardiner2002}, where 
\begin{equation}
\label{eq:probinteval}
G(T) = \int_a^b p(x,T)\, dx
\end{equation}
is the probability that the particle is at time $T$ in $(a,b)$ when it has
started at zero time at $x_0$ and we have calculated the PDF $p(x,t)$ imposing
absorbing boundary conditions on those boundaries where the particle can leave
the interval. Writing the Fokker-Planck, or forward Kolmogorov, equation in the
form $\partial_t p + \partial_x j = 0$ we have readily
\begin{equation}
\label{eq:Gdot}
\dot{G}(T) = -j(x,T) \Big|_{a}^{b}.
\end{equation}
If the upper boundary is a natural boundary at $\infty$, $j(\infty,T) = 0$,
and we are interested in hits at the origin we have $ f(T) = -j(0,T)$. For our
problem the probability current density at the origin  has been calculated in
Sec.~\ref{sec:heurclass}. From Eqs.~(\ref{eq:asymptoticcurrent}) and
(\ref{eq:explAbsBCsolution}) one obtains
\begin{equation}
f(T) = \frac{1}{\Gamma(\nu)} \left(\frac{x_0^2}{4D}\right)^\nu
T^{-(\nu+1)}\exp\left(-\frac{x_0^2}{4DT}\right).
\end{equation}
For long times this is a power law $f(T) \propto T^{-(3-n)/2}$ \cite{Bray2000}.

The result was obtained solving the Fokker-Planck equation
(\ref{eq:Fokker-Planck}) on the semi-infinite interval $(0,\infty)$ with an
absorbing boundary condition at $x = 0$ and the initial condition at $x = x_0$,
and it is restricted to $n < 1$, i.e.\ $\nu > 0$. The spectrum of this boundary
value problem is continuous.

However, if we are interested in the first time to leave a finite interval,
we have to solve a boundary value problem with, for example, absorbing boundary
conditions at both ends of the interval which typically has a discrete
spectrum. We find it preferable to adopt a more formal approach, based on the
backward Kolmogorov equation. The boundary value problem can then be
transformed to a canonical Sturm-Liouville problem and systematically solved. 

\subsection{The backward Kolmogorov equation}

In this section we use a special Fokker-Planck technique proposed by Kearney
and Majumdar \cite{Kearney2005} to obtain a differential equation for the
first-passage time PDF in Laplace space. Their method is very powerful, because
the boundary conditions can be easily established in Laplace space and the
functional $V[X_t]$ can be chosen such that different relevant quantities can
be computed. Therefore we present the application of this method to our problem
in some detail.

Considering a stochastic process starting at $X_0 = x$ governed by the
stochastic differential equation~(\ref{eq:sde}), we are interested in the PDF
$f(T_b,x)$ of the first-passage time $T_b$ with respect to a certain level $b$,
i.e.\ the time when the process has reached the level $b$ for the first time.
First of all we define an arbitrary functional $V[X_t]$ by
\begin{equation}
\label{eq:functional1}
T = \int_0^{T_b} V[X_t]\, dt.
\end{equation}
$T$ can have several meanings; in the special case $V[X_t] \equiv 1$ it is
simply the first-passage time $T_b$. The strategy is to find a differential
equation in Laplace space for $f(T,x)$. The Laplace transform of $f(T,x)$ with
respect to $T$ is given by
\begin{eqnarray}
\label{eq:Laplace1}
\tilde{f}(s,x) & = & \mathcal{L}_T [f(T,x)](s) \nonumber \\
& = & \int_0^\infty f(T,x) e^{-sT}\, dT = \langle e^{-sT} \rangle_T,
\end{eqnarray}
where $s \in \mathbb{C}$. Splitting the interval $[0,T_b]$ into a small
interval $[0,\Delta t]$ and an interval $(\Delta t,T_b]$, we can expand the
integral over the small interval to first order in $\Delta t$:
\begin{equation}
\label{eq:approx1}
\int_0^{\Delta t} V[X_t]\, dt = V[x]\Delta t + o(\Delta t).
\end{equation}
Thus Eq.~(\ref{eq:functional1}) becomes
\begin{equation}
\label{eq:functional2}
T = V[x]\Delta t + \int_{\Delta t}^{T_b} V[X_t]\, dt =: T_1 + T_2.
\end{equation}
Inserting Eq.~(\ref{eq:functional2}) into Eq.~(\ref{eq:Laplace1}) gives
\begin{eqnarray}
\label{eq:Laplace2}
\tilde{f}(s,x) & = & \langle e^{-sT} \rangle_T
                 = \langle e^{-sT_1} e^{-sT_2} \rangle_T \nonumber \\
               & = & \int_0^\infty f(T,x) e^{-sT_1} e^{-sT_2}\, dT.
\end{eqnarray}
If we split the interval $[0,T_b]$ as described above, we must take into
account that we also split our trajectory in two, where the starting point of
the second part, $y := X_{\Delta t} = x + \Delta x$, is random itself.
Therefore the PDF takes the form
\begin{eqnarray}
\label{eq:density}
f(T,x) & = & \int_0^b f(T_1,x) f(T_2,y)\, dy \nonumber \\
       & = & \int_{-x}^{b-x} f(T_1,x) f(T_2,x+\Delta x)\, d(\Delta x).
\end{eqnarray}
Inserting this into Eq.~(\ref{eq:Laplace2}) and taking into account that $T_1$
is constant, and hence $dT = dT_2$, we obtain
\begin{equation}
\label{eq:Laplace3}
\tilde{f}(s,x)
= e^{-sV[x]\Delta t} \langle \tilde{f}(s,x+\Delta x) \rangle_{\Delta x},
\end{equation}
where the average is done over all realizations of $\Delta x$. With Taylor
expansions around $x$ of $e^{-sV[x]\Delta t}$ to the first order and of
$\tilde{f}(s,x+\Delta x)$ to the second order, Eq.~(\ref{eq:Laplace3}) becomes
\begin{multline}
\label{eq:approx3}
\tilde f(s,x) = \left(1-sV[x]\Delta t\right) \\
\times \left[ \tilde f(s,x)+ \frac{\partial\tilde f(s,x)}{\partial x} \langle
\Delta x \rangle + \frac{1}{2}\frac{\partial^2\tilde{f}(s,x)}{\partial x^2}
\langle \Delta x^2 \rangle \right].
\end{multline}
In a first order approach
\begin{equation}
\label{eq:Langevin2}
\Delta x = \frac{nD}{x}\Delta t + \sqrt{2D}\Delta W_t,
\end{equation}
where $\Delta W_t = W_{t+\Delta t} - W_t$, and thus, using Eq.~(\ref{eq:mean}),
\begin{equation}
\langle \Delta x \rangle = \frac{nD}{x} \Delta t.
\end{equation}
Then the mean value of $\langle \Delta x^2 \rangle$ is, making again use of the
zero mean property of the Wiener process, as well as of its autocorrelation
function given in Eq.~(\ref{eq:autocorrelation}),
\begin{equation}
\langle \Delta x^2 \rangle = 2D \Delta t + o(\Delta t).
\end{equation}
Finally, putting $V[X_t] \equiv 1$, we get the desired backward Kolmogorov
equation for the first passage time PDF in Laplace space:
\begin{equation}
\label{eq:Kearney}
\frac{\partial^2 \tilde{f}(s,x)}{\partial x^2}
+ \frac{n}{x}\frac{\partial \tilde{f}(s,x)}{\partial x}
- \frac{s}{D}\tilde{f}(s,x) = 0.
\end{equation}

\subsection{Formulation of the boundary value problem}

We now proceed to the formulation of the boundary value problems corresponding
to the solutions of the first-passage time PDFs, distinguishing between the
three classes of boundaries the origin can belong to, as discussed in
Sec.~\ref{sec:classification_boundaries}. On the right side we impose an
absorbing boundary at $b$: the first-passage time vanishes for $x\to b^-$,
i.e., $f(T, x\to b^-)=\delta (T)$. Inserting this into Eq.~(\ref{eq:Laplace1})
gives
\begin{equation}
\label{eq:boundary_a}
\lim_{x \to b^-} \tilde f(s,x) = 1.
\end{equation}

The simplest case is if the zero level is an entrance boundary, i.e.\ $n \geq
1$. Starting from an inititial value $X_0 = x > 0$, the zero level can never be
reached, which corresponds to a reflecting wall at the origin. Applying
standard arguments for reflecting boundaries \cite{Gardiner2002}, the
corresponding boundary condition is
\begin{equation}
\label{eq:boundary_0_ref}
\lim_{x \to 0^+} \frac{\partial \tilde{f}(s,x)}{\partial x} = 0.
\end{equation}

For $n \leq -1$ the origin is an exit boundary. This means that it is
impossible to reach any interior point of the state space if the initial
point approaches the origin. This means that we have an absorbing boundary
corresponding to
\begin{equation}
\label{eq:boundary_0_abs}
\lim_{x \to 0^+} \tilde{f}(s,x) = 1,
\end{equation}
and the first-passage time with respect to $x = b$ will diverge. Instead of
the first-passage time the analysis of the previous section resulting in the
backward Kolmogorov equation~(\ref{eq:Kearney}) together with the boundary
conditions~(\ref{eq:boundary_a}) and (\ref{eq:boundary_0_abs}) gives the
first-exit time from the interval $(0,b)$. 

In the case of a regular boundary, which
happens for $-1< n < 1$, the behaviour is the most complicated. The process can
both reach and leave the boundary zero, which means that also zero crossings are
possible and the support of the process is the whole real axis. The first-exit
time from $(0,b)$ is again given by the same boundary condition problem as in
the case of the exit boundary.

For the sake of simplicity we rename $\tilde{f}(s,x) =: y(x)$. Restricting the
process to the positive half axis, our boundary value problem for the three
kinds of boundary in the origin reads
\begin{gather}
\label{eq:boundary_value_problem}
y''(x) + \frac{n}{x} y'(x) - \frac{s}{D} y(x) = 0,\\
\label{eq:boundary_conditions}
\mathbf{A}\mathbf{y}(0) + \mathbf{B}\mathbf{y}(a) = \mathbf{c},
\end{gather}
where
\begin{equation}
\setcounter{MaxMatrixCols}{2}
\mathbf{y}(x) = \begin{pmatrix} y(x) \\ y'(x) \end{pmatrix},\quad
\mathbf{B} = \begin{pmatrix} 0 & 0\\ 1 & 0 \end{pmatrix}.
\end{equation}
An absorbing boundary at zero corresponds to
\begin{equation}
\setcounter{MaxMatrixCols}{2}
\mathbf{A} = \begin{pmatrix} 1 & 0\\ 0 & 0 \end{pmatrix},\quad
\mathbf{c} = \begin{pmatrix} 1 \\ 1 \end{pmatrix},
\end{equation}
whereas a reflecting boundary at zero corresponds to
\begin{equation}
\mathbf{A} = \begin{pmatrix} 0 & 1\\ 0 & 0 \end{pmatrix},\quad
\mathbf{c} = \begin{pmatrix} 0 \\ 1 \end{pmatrix}.
\end{equation}
Multiplying Eq.~(\ref{eq:boundary_value_problem}) with the integrating factor
$\exp\left(\int \frac{n}{x} dx\right)$ leads to
\begin{equation}
\label{eq:Sturm-Liouville1}
-(x^n y')' = -\frac{s}{D} x^n y.
\end{equation}
This is the canonical Sturm-Liouville form \cite{Zettl2005}
\begin{equation}
\label{eq:Sturm-Liouville2}
-(py')'+ qy = \lambda w y,
\end{equation}
with $p(x) = x^n$, the weighting function $w(x) = x^n$, $q(x) \equiv 0$,
and the spectral parameter $\lambda = -s/D$.

We now observe that $u := y - 1$ transforms the homogeneous problem
(\ref{eq:boundary_value_problem}) with inhomogeneous boundary conditions
(\ref{eq:boundary_conditions}) into an inhomogeneous problem with homogeneous
boundary conditions
\begin{eqnarray}
\label{eq:inhomogeneous_Sturm-Liouville1}
& &-(pu')'  =  \lambda wu + \lambda w,\\
\label{eq:Dirichlet_Neumann}
& & \mathbf{A}\mathbf{u}(0) + \mathbf{B}\mathbf{u}(a)  =  0,
\end{eqnarray}
where $\mathbf{u}(x) = \left( u(x),\ u'(x) \right)^\mathsf{T}$, and the two
possible choices of $\mathbf{A}$ and $\mathbf{c}$ correspond to the Dirichlet
problem and the Dirichlet-Neumann problem, respectively. This is easier to
solve, since it determines a self-adjoint operator $\mathcal{L}$ defined by
\begin{equation}
\label{eq:operator}
\mathcal{L} u = \frac{1}{w} \left[-(pu')'\right]
\end{equation}
in the weighted Hilbert space $H = L^2(J,w)$, where we have defined the open
interval $J = (0,b)$. This operator is not to be confused with the Laplace
transformation operator $\mathcal{L}_T$ in Eq.~(\ref{eq:Laplace1}), which can be recognized
from the index indicating the transformed variable.

This can be seen as follows. Let $u,v \in H$; then the inner product is given
by $\langle u,v \rangle = \int_0^b \bar{u} v w\, dx$; taking into account that
$u$ and $v$ satisfy the homogeneous boundary conditions
(\ref{eq:Dirichlet_Neumann}), we get after integrating twice by parts
\begin{eqnarray}
\langle u, \mathcal{L}v \rangle & = & -\int_0^b\bar{u}(pv')'\,dx \nonumber \\
& = & \Big[p(v\bar{u}'-\bar{u}v')\Big]_0^b -\int_0^b(p\bar{u}')'v\,dx\nonumber\\
& = & \langle \mathcal{L}u,v \rangle.
\end{eqnarray}
Using the definition from Eq.~(\ref{eq:operator}) the boundary value
problem given by Eqs.~(\ref{eq:boundary_value_problem}--\ref{eq:boundary_conditions})
can be simplified to
\begin{eqnarray}
\label{eq:inhomogeneous_Sturm-Liouville2}
(\mathcal{L} -\lambda\mathbf{1}) u & = & \lambda, \\
\label{eq:Dirichlet-Neumann2}
\mathbf{A}\mathbf{u}(0) + \mathbf{B}\mathbf{u}(b) & = & 0.
\end{eqnarray}

\subsection{Formal solution of the boundary value problem}


We now exploit the property that the homogeneous boundary value problem with
homogeneous boundary conditions
\begin{eqnarray}
(\mathcal{L} -\alpha\mathbf{1}) u & = & 0, \label{eq:homogeneous_Sturm-Liouville} \\
\mathbf{A}\mathbf{u}(0) + \mathbf{B}\mathbf{u}(b) & = & 0 ,
\end{eqnarray}
has nontrivial solutions $u_k$ with eigenvalues $\alpha_k,\ k \in \mathbb{N}$,
\begin{equation}
\label{eq:homogeneous_Sturm-Liouville1}
\mathcal{L} u_k = \alpha_k u_k.
\end{equation}
Because $\mathcal{L}$ is self-adjoint, the eigenvalues $\alpha_k$ are real
and the eigenfunctions $u_k$ form an orthonormal basis of $H$. Furthermore
$\alpha_k > 0$ holds, since $\alpha_k = \langle u_k, \mathcal{L} u_k \rangle$.
Hence the solution $u$ of the inhomogeneous problem given by
Eqs.~(\ref{eq:inhomogeneous_Sturm-Liouville2}) and
(\ref{eq:Dirichlet-Neumann2}) can be expressed through an expansion in this
basis,
\begin{equation}
\label{eq:Fourier}
u = \sum_{k=1}^\infty c_k u_k,
\end{equation}
with $c_k = \langle u_k, u \rangle$. Inserting $u=1$ gives the normalization, 
$\sum_{k=1}^\infty \langle u_k, 1 \rangle u_k = 1$. The coefficients $c_k$ can 
be derived from Eq.~(\ref{eq:inhomogeneous_Sturm-Liouville2}):
\begin{equation}
\langle u_k, \mathcal{L} u \rangle - \langle u_k, \lambda u \rangle
= \langle u_k, \lambda \rangle.
\end{equation}
Again, making use of the definition of a self-adjoint operator, we can pull
$\mathcal{L}$ into the first component of the inner product. Employing
Eq.~(\ref{eq:homogeneous_Sturm-Liouville1}) we get
\begin{equation}
c_k = \frac{\langle u_k, \lambda \rangle}{\alpha_k -\lambda}.
\end{equation}
The solution of the inhomogeneous problem reads
\begin{equation}
u = \sum_{k=1}^\infty \frac{\langle u_k,\lambda \rangle}{\alpha_k-\lambda} u_k. 
\end{equation}
Because the eigenfunctions $u_k$ do not depend on $\lambda = -s/D$ and the
Laplace transformation is a linear operation we obtain the inverse Laplace
transform of $y = 1 + u$ as
\begin{eqnarray}
y(T,x) & = & \mathcal{L}_s^{-1}[y(s,x)](T) \nonumber \\
& = & \mathcal{L}_s^{-1}[1] + \sum_{k=1}^\infty \langle u_k, 1 \rangle u_k\,
\mathcal{L}_s^{-1}\left[\frac{\lambda}{\alpha_k -\lambda}\right] \\
& = &\!\delta(T)\!  +\! \! \sum_{k=1}^\infty \langle u_k, 1 \rangle u_k
[\alpha_kD e^{-\alpha_kDT}\! - \! \delta(T)].\nonumber 
\end{eqnarray}
Since the $u_k$ are normalized the two delta functions cancel out. Returning to
our original notation, we write the final result for the first-passage time (or
first-exit time when appropriate) PDF as 
\begin{equation}
\label{eq:solution}
f(T,x) = \sum_{k=1}^\infty \langle u_k, 1 \rangle u_k
\alpha_kD e^{-\alpha_kDT}.
\end{equation}
Of course, this PDF is normalized to 1: knowing that $\alpha_k > 0$ we have
\begin{eqnarray}
\label{eq:normalization}
\int_0^\infty f(T,x)\, dT & = & \sum_{k=1}^\infty \langle u_k, 1 \rangle u_k
\alpha_kD \int_0^\infty e^{-\alpha_kDT}\, dT \nonumber \\
& = & \sum_{k=1}^\infty \langle u_k, 1 \rangle u_k = 1.
\end{eqnarray}
We are now able to solve the specific boundary value problems for the three
different kinds of boundaries at zero.

\subsection{Comparison of theory and simulation}

\subsubsection{Simulation method}
\label{sec:simulation}

To simulate the process $X_t$ that fulfills Eq.~(\ref{eq:sde}), we have used
the Euler-Maruyama method \cite{Maruyama1955,Milstein1995,Kloeden1999,
Higham2001}, which in this case with an additive noise is identical to the
higher-order Milstein method \cite{Milstein1995,Kloeden1999,Higham2001}.
A generic autonomous stochastic differential equation
\begin{equation}
\label{eq:sde_generic}
dX_t = \mu(X_t) dt + \sigma(X_t) dW_t
\end{equation}
can be integrated between two successive times $t_n$ and $t_{n+1}$, giving
\begin{equation}
X_{n+1} = X_n + \int_{t_n}^{t_{n+1}} \mu(X_t) \, dt
+ \int_{t_n}^{t_{n+1}} \sigma(X_t) \, dW_t,
\end{equation}
where $X_n$ is short for $X_{t_n}$.
The approximation of the integrands to their value in $t_n$,
\begin{eqnarray}
\mu(X_t) &\approx& \mu(X_n), \nonumber \\
\sigma(X_t) &\approx& \sigma(X_n),
\end{eqnarray}
yields the Euler method for the It\=o case, called Euler-Maruyama
\cite{Maruyama1955},
\begin{equation}
\label{eq:Euler-Maruyama}
X_{n+1} = X_n + \mu(X_n) \Delta t + \sigma(X_n) \Delta W_n,
\end{equation}
where $\Delta t = t_{n+1}-t_n$ and $\Delta W_n = W_{n+1}-W_n \sim N(0,\Delta t)
\sim \sqrt{\Delta t}N(0,1)$, i.e.\ $\Delta W_n$ it is a normal random
variable with PDF
\begin{equation}
p(w) = \frac{1}{\sqrt{2\pi\Delta t}}\exp\left(-\frac{w^2}{2\Delta t}\right).
\end{equation}
The Euler-Maruyama method has strong order of convergence 1/2.
The Milstein method raises this to 1 adding to the right-hand side of
Eq.~(\ref{eq:Euler-Maruyama}) the correction $\frac{1}{2}\sigma(X_n)
\sigma'(X_n)[(\Delta W_n)^2-\Delta t]$, where $\sigma'(X_n) = d\sigma(x)/dx
|_{x=X_n}$. However, for an additive noise this derivative vanishes and so here
the correction is zero. Schemes with order higher than 1 contain further terms
some of which are nonzero also for additive noise, though many cancel out with
respect to the general case, called multiplicative \cite{Fox1972}, where
$\sigma$ depends on $X_t$.

The approximation of the noise term as
\begin{equation}
\label{eq:stratonovich1}
\sigma(X_t) \approx \frac{\sigma(X_n)+\sigma(X_{n+1})}{2}
\end{equation}
or as
\begin{equation}
\label{eq:stratonovich2}
\sigma(X_t) \approx \sigma\left(\frac{X_n+X_{n+1}}{2}\right)
\end{equation}
yields the corresponding method for the Stratonovich case; if $\sigma$ is
continuous, both Eqs.~(\ref{eq:stratonovich1}) and (\ref{eq:stratonovich2})
lead to the same limit for $\Delta t \to 0$. This results in an implicit
method, where to compute $X_{n+1}$ it is required to estimate it before; the
predictor-corrector approach where $X_{n+1}$ in Eq.~(\ref{eq:stratonovich1}) or
(\ref{eq:stratonovich2}) is approximated by Eq.~(\ref{eq:Euler-Maruyama}) for
the It\=o case is known by the name of Euler-Heun or Heun \cite{Milstein1995,
Kloeden1999}. As already observed at the beginning of Sec.~\ref{sec:heurclass}
with respect to the Fokker-Planck equation, both the It\=o and the Stratonovich
convention lead to the same result when the noise is additive as here.
Interestingly the Milstein scheme represents both the order 1 strong
It\=o-Taylor approximation and the order 1 strong Stratonovich-Taylor
approximation, i.e.\ even in the multiplicative case it coincides for both
kinds of stochastic integral.

In other words, the choice of $X_t$ within the discretization interval
$[X_n,X_{n+1}]$ affects the outcome of the integration only as far as the
dependence of the noise term $\sigma$ on $X_t$ is concerned, because the
covariation of $X_t$ and of the Wiener process $W_t$ driving the stochastic
integral is not zero, $[X_t,W_t] \neq 0$ \cite{Germano2009} (unfortunately
closed intervals and covariations share the same notation). The choice of $X_t
\in [X_n,X_{n+1}]$ has no influence on the integration of the drift term $\mu$
with respect to $t$, and the choice of $t \in [t_n,t_{n+1}]$ does not matter
for either $\mu$ or $\sigma$ if they depend on $t$, as $[t,t] = 0$ and $[t,W_t]
= 0$; in the three latter cases the same limit results for $\Delta t \to 0$.

Eq.~(\ref{eq:Euler-Maruyama}) can be implemented straightforwardly in code.
However, measuring the first-passage time with respect to a certain level needs
a further refinement, since there is a finite hitting probability during each
discretized time interval $\Delta t$, and thus the first-passage time is
overestimated. An analytic expression for the probability that the process hits
the level $b$ during a discretization interval $\Delta t$ was found by Mannella
\cite{Mannella1999}. If we introduce the abreviations $\mu_n = \mu(X_n)$,
$\mu_b = \mu(b)$ and $\mu'_b = d\mu(x)/dx |_{x=b}$, the hitting probability
reads
\begin{multline}
P(\textrm{hit}) = \exp\left\{-\frac{\mu'_b}{2D\left(e^{2\mu'_b\Delta t} - 1
\right)} \right.\\
\left.\times\left[X_{n+1} -b+(X_n -b)e^{\mu'_b\Delta t}
- \frac{\mu_b}{\mu'_b}\right]^2\right. \\
+ \left. \frac{1}{4D\Delta t} \left[X_{n+1} - \left( X_n +
\frac{\mu_n + \mu_{n+1}}{2} \Delta t\right) \right]^2 \right\}.
\end{multline}

We can now summarize the simulation algorithm. We draw a Gaussian random number
$\Delta W_n$ using e.g.\ the Box-Muller method \cite{Box1958} and propagate the
process $X_n$ by a time step $\Delta t$. If the propagated value exceeds the
level $b$ for the first time, i.e.\ $X_{n+1} > b$, the process is terminated.
Otherwise we check for missed hits in the discretization interval by drawing a
uniformly distributed random number $U \in [0,1)$ and accepting the hitting
hypothesis if $P(\textrm{hit}) > U$; this fulfills the second terminating
condition. In both cases the first-passage time is set to $t_n$, i.e.\ the
value before the propagation.

In the case of an entrance boundary a further refinement of the simulation
algorithm is possible. Knowing that the zero level can never be reached from
the interior of the state space of the process, it is clear that negative
values in the simulations must result from discretization errors. If this is
the case we can reduce the time step until the propagated value of the process
is positive.

\subsubsection{Entrance boundary}

As we know from the classification of the origin, we have an entrance boundary
for $n \ge 1$ (i.e. $\nu \le 0$). The general solution of the homogeneous
differential equation~(\ref{eq:homogeneous_Sturm-Liouville}) is
\begin{equation}
\label{eq:generalsolution1}
u(x) = x^\nu \left[ A J_\nu \left( \sqrt{\alpha} x \right) + B  Y_\nu
\left( \sqrt{\alpha} x \right)\right],
\end{equation}
where $J_\nu$ and $Y_\nu$ are the Bessel functions of the first and second
kind, respectively. 

Exploiting $J_\nu'(x)=J_{\nu-1}(x) -(\nu/x)J_\nu(x)$ and an analogous
formula for $Y_\nu$ \cite{Arfken1985}, we obtain the derivative
\begin{equation}
\label{eq:generalsolution1derivative}
u'(x) = \sqrt{\alpha} \, x^\nu \left[ A J_{\nu -1} \left( \sqrt{\alpha} x
\right) + B Y_{\nu -1} \left( \sqrt{\alpha} x \right) \right].
\end{equation}

The relevant boundary conditions for $\tilde f(s,x)$ given by
Eqs.~(\ref{eq:boundary_a}) and (\ref{eq:boundary_0_ref}) transform to
$u(b) = 0$ (absorption at $x = b$) and  $\lim_{x \to 0} u'(x) = 0$
(reflection at $x = 0$), respectively.

To evaluate the eigenfunctions in the case of negative and integer $\nu$ one
can use the symmetry relation \cite{Arfken1985} 
\begin{equation}
J_{-\nu}(z) = (-1)^\nu J_\nu(z),
\end{equation}
which holds for integer $\nu$, to see that the first term in 
Eq.~(\ref{eq:generalsolution1derivative}) goes to zero for $x\to 0$, 
because its leading order term behaves as $x$. Since the Bessel functions of
second kind diverge as $x \to 0$, the reflecting boundary condition can be
fulfilled only if $B = 0$.

The absorbing boundary condition at $x = b$ determines the eigenvalues of the
problem by the requirement that $J_\nu(\sqrt{\alpha_k}b) = 0$. Denoting the
$k$th zero of $J_\nu(x)$ by $j_k$ we thus have $u_k(x)=A_k\,x^\nu J_\nu\left(
j_k {x}/{b}\right)$. The constant $A_k$ is determined by the condition $\langle
u_k,u_l \rangle = \delta_{kl}$. Remember that the brackets denote the scalar
product in the weighted Hilbert space with weigth $w = x^n$. Observing the
orthogonality relation
\cite{Watson1962}
\begin{equation}
\label{eq:Bessel_orthogonality}
\int_0^b J_\nu\left(j_k\frac{x}{b}\right)
         J_\nu\left(j_l\frac{x}{b}\right) x\, dx
= \frac{1}{2}\, b^2 J_{\nu+1}^2(j_k)\; \delta_{kl}
\end{equation}
one obtains $A_k= \sqrt{2}\,b^{-1}\,/J_{\nu+1}(j_k)$, so that 
\begin{equation}\label{eq:eigenfunctions1}
u_k(x)=\sqrt{2}\;b^{-1}\;x^\nu J_\nu\left(j_k \frac{x}{b}\right)/J_{\nu+1}(j_k).
\end{equation}
We can further compute \cite{Gradshteyn1965}
\begin{eqnarray}
\label{eq:scalarproducts1}
\langle u_k, 1 \rangle = \frac{\sqrt{2}\,b^{1-\nu}}{j_k}
\left[\frac{(j_k/2)^{\nu -1}}{\Gamma(\nu)J_{\nu +1}(j_k)}
- \frac{J_{\nu-1}(j_k)}{J_{\nu+1}(j_k)} \right].
\end{eqnarray}
For integer $\nu$ a recurrence relation $J_{\nu+1}(x) + J_{\nu-1}(x)
= 2\nu J_\nu(x)/x$ holds, which, evaluated at the $k$th zero of $J_\nu$,
delivers $J_{\nu-1}(j_k)=-J_{\nu+1}(j_k)$. Hence Eq.~(\ref{eq:scalarproducts1})
simplifies to
\begin{eqnarray}
\label{eq:scalarproducts1integer}
\langle u_k, 1 \rangle = \frac{\sqrt{2}\,b^{1-\nu}}{j_k}
\left[ \frac{(j_k/2)^{\nu-1}}{\Gamma(\nu)J_{\nu +1}(j_k)} +1 \right].
\end{eqnarray}

For non-integer $\nu$ the Bessel functions $J_\nu$ and $J_{-\nu}$ are two
linear independent solutions of Eq.~(\ref{eq:homogeneous_Sturm-Liouville}),
and it is more convenient to write the general solution as 
\begin{equation}
\label{eq:generalsolution2}
u(x) = x^\nu \left[ A J_\nu \left( \sqrt{\alpha} x \right) + B J_{-\nu}
\left( \sqrt{\alpha} x \right) \right].
\end{equation}

Exploiting $J_{-\nu}'(x) = -J_{1-\nu}(x) -(\nu/x)J_{-\nu}(x)$
\cite{Arfken1985}, the derivative can be written as
\begin{equation}\label{eq:u_deriv_noninteger_nu}
u'(x) = \sqrt{\alpha} \, x^\nu \left[AJ_{\nu-1}(\sqrt{\alpha}x)
- B J_{1-\nu}(\sqrt{\alpha}x)\right].
\end{equation}

In this case $x^\nu J_{\nu-1}(\sqrt{\alpha}x)$ diverges as $x \to 0$, and
the left boundary condition requires  $A = 0$. The right boundary condition
determines the eigenvalues similarly as in the previous case; it is required
that $\sqrt{\alpha}b$ are the zeros $j_k$ of the Bessel function $J_{-\nu}$.
Again the second constant is evaluated using
Eq.~(\ref{eq:Bessel_orthogonality}). The normalized eigenfunctions are
\begin{equation}
\label{eq:eigenfunctions2}
u_k(x) = \sqrt{2}\; b^{-1} x^\nu J_{-\nu} \left(j_k
\frac{x}{b}\right)/{J_{1-\nu}(j_k)}
\end{equation}
with
\begin{equation}
\label{eq:scalarproducts2}
\langle u_k, 1 \rangle = \sqrt{2}\; b^{1-\nu} j_k^{-1}.
\end{equation}



For completeness we prove that the eigenfunctions given by
Eqs.~(\ref{eq:eigenfunctions1}) and (\ref{eq:eigenfunctions2}) fulfill the
normalization condition~(\ref{eq:normalization}). Setting $x/b = z$,
for integer $\nu$ Eqs.~(\ref{eq:eigenfunctions1}) and
(\ref{eq:scalarproducts1integer}) yield
\begin{multline}
\sum_{k=1}^\infty \langle u_k, 1 \rangle u_k
= z^\nu \sum_{k=1}^\infty \frac{2J_\nu(j_kz)}{j_k J_{\nu+1}(j_k)}
\left[ \frac{(2/j_k)^{1-\nu}}{\Gamma(\nu)J_{\nu+1}(j_k)} + 1 \right] \\
= z^\nu \sum_{k=1}^\infty \left[ \frac{(2/j_k)^{2-\nu} J_\nu(j_kz)}
{\Gamma(\nu)J_{\nu+1}^2(j_k)} + \frac{2J_\nu(j_kz)}{j_kJ_{\nu+1}(j_k)} \right]
= 1,
\end{multline}
where we have used the Fourier-Bessel expansions
\begin{gather}
z^\nu = \sum_{k=1}^\infty \frac{2J_\nu(j_kz)}{j_kJ_{\nu+1}(j_k)}
\label{eq:Fourier-Bessel1} \\
z^{-\nu} - z^\nu = \sum_{k=1}^\infty 
\frac{(2/j_k)^{2-\nu}J_\nu(j_kz)}{\Gamma(\nu)J_{\nu+1}^2(j_k)}.
\label{eq:Fourier-Bessel2}
\end{gather}
Eq.~(\ref{eq:Fourier-Bessel1}) is found in Watson \cite{Watson1962},
Eq.~(\ref{eq:Fourier-Bessel2}) is proved in the appendix.
For non-integer $\nu$ Eqs.~(\ref{eq:eigenfunctions2}),
(\ref{eq:scalarproducts2}) and (\ref{eq:Fourier-Bessel1}) yield
\begin{equation}
\sum_{k=1}^\infty \langle u_k, 1 \rangle u_k = z^\nu
\sum_{k=1}^\infty \frac{2J_{-\nu}(j_kz)}{j_kJ_{1-\nu}(j_k)} = 1.
\end{equation}

\begin{figure}[htbp]
\includegraphics[width=0.49\columnwidth]{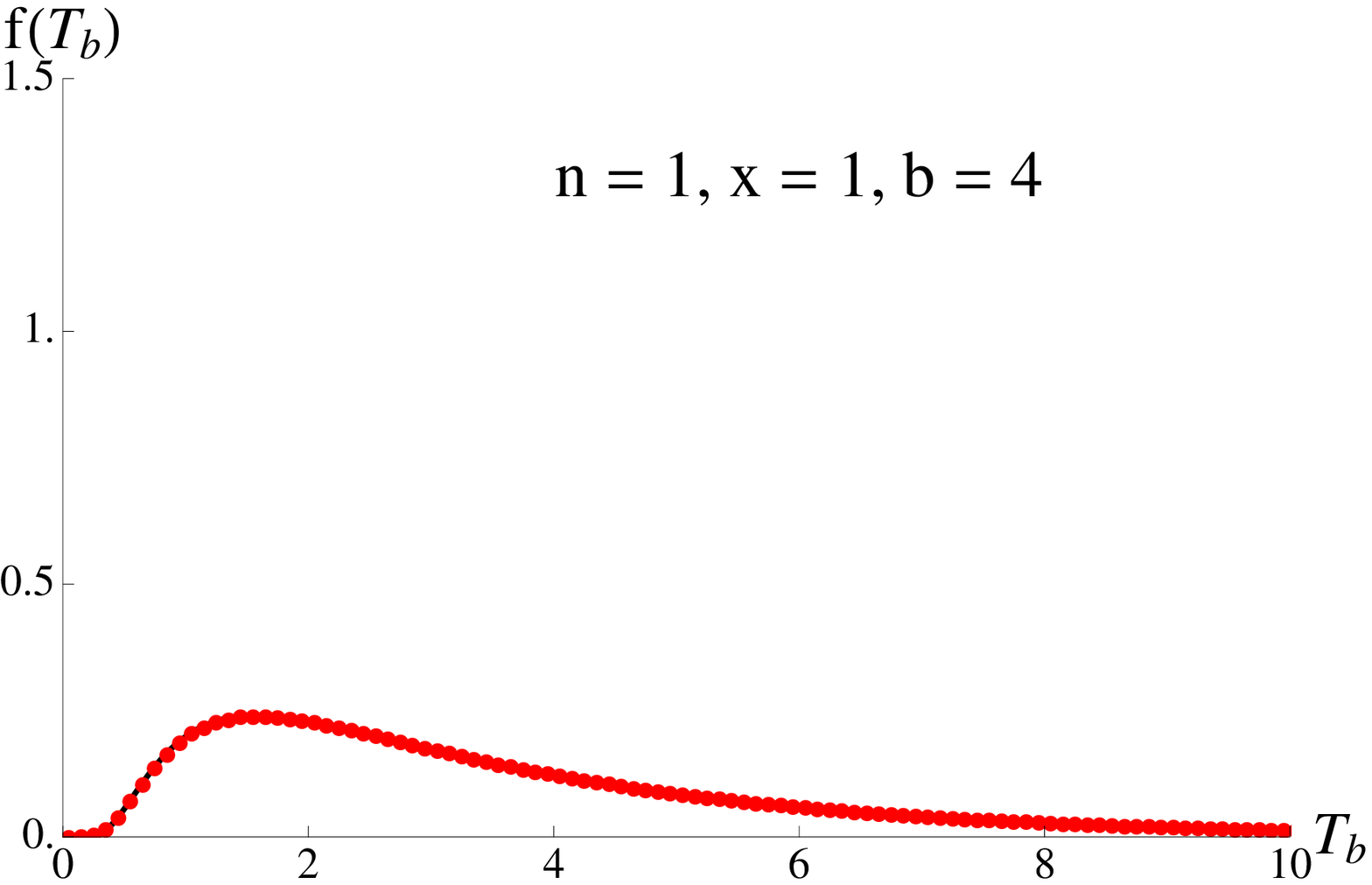}
\includegraphics[width=0.49\columnwidth]{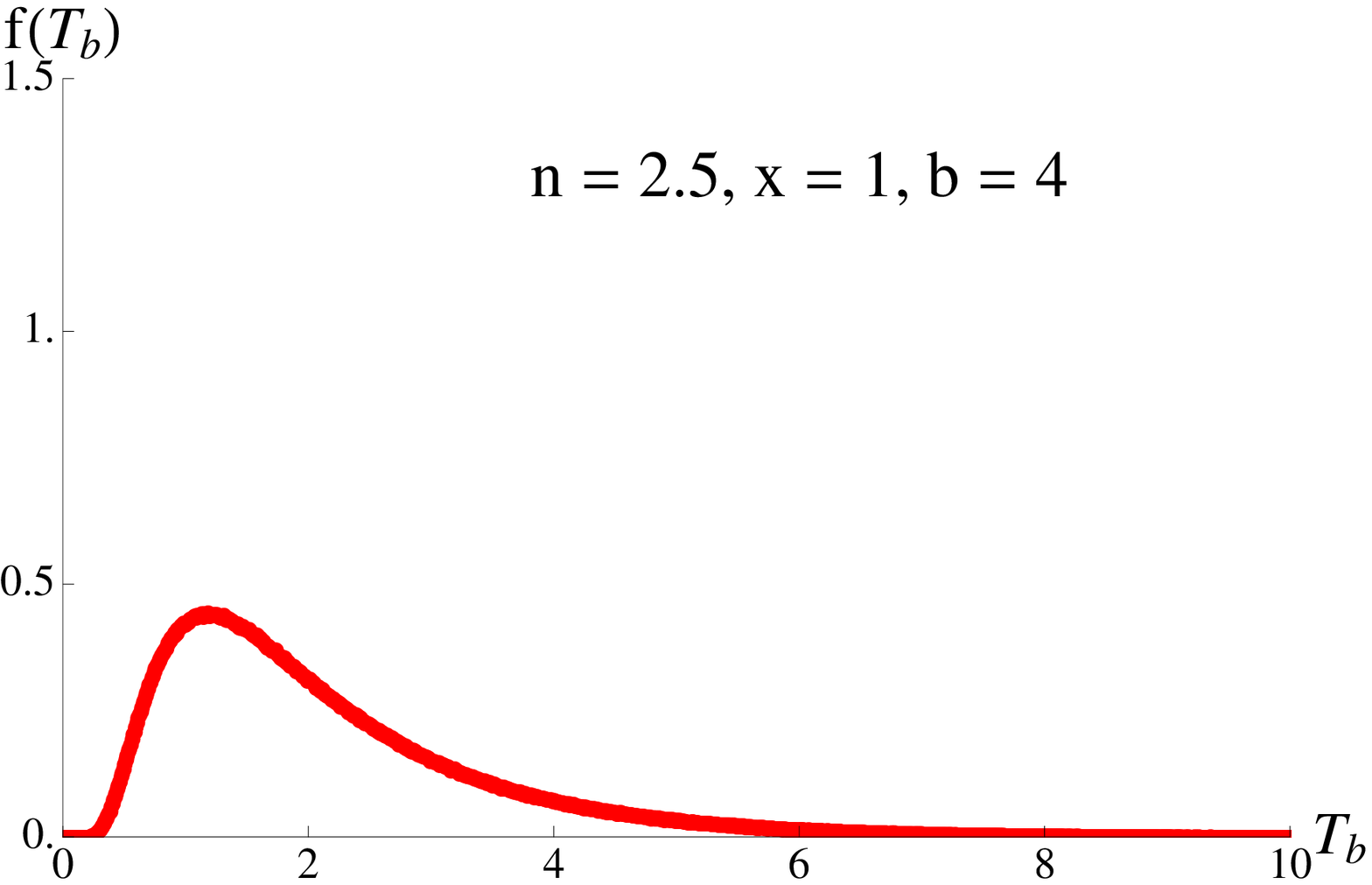}
\includegraphics[width=0.49\columnwidth]{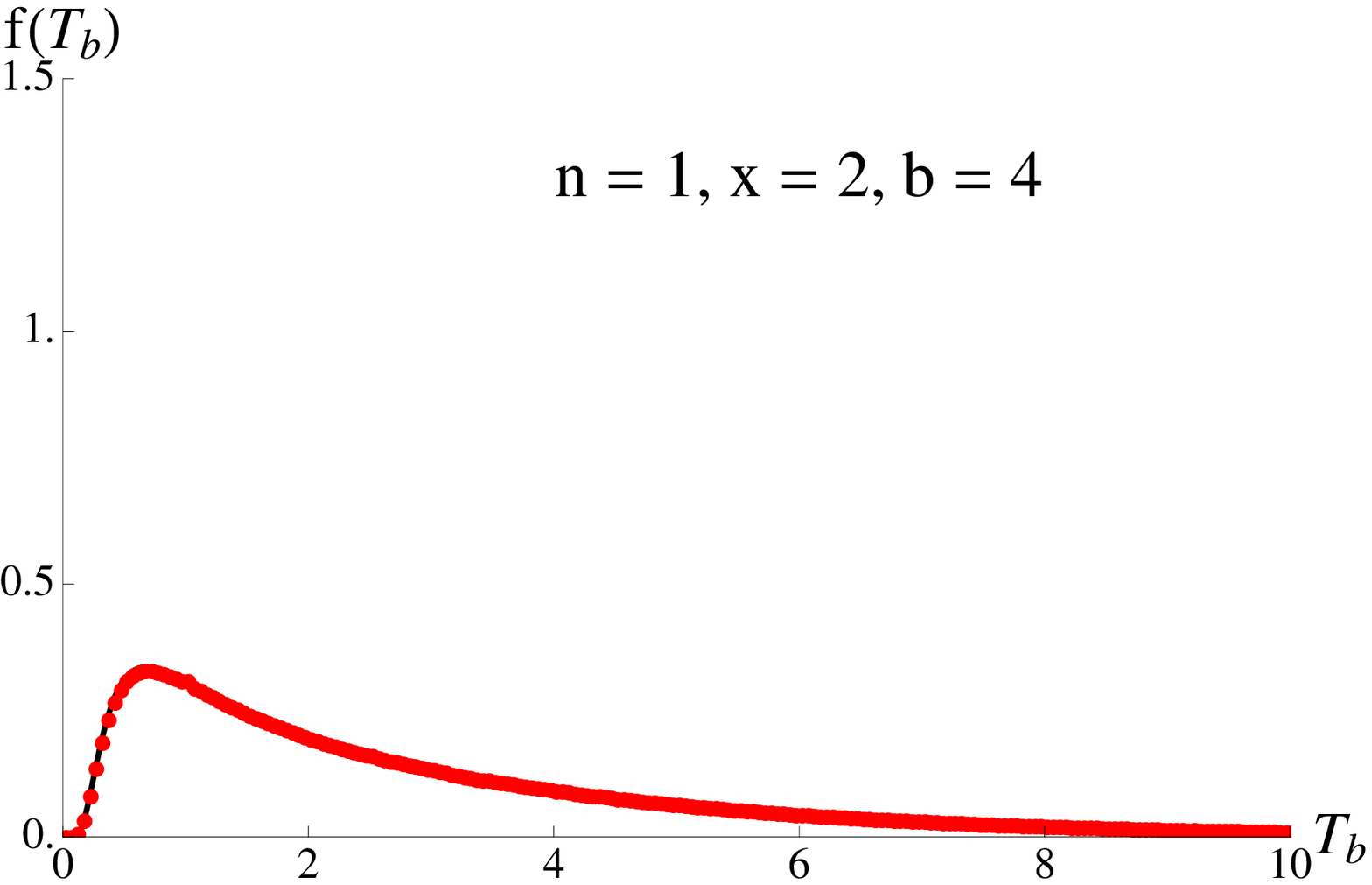}
\includegraphics[width=0.49\columnwidth]{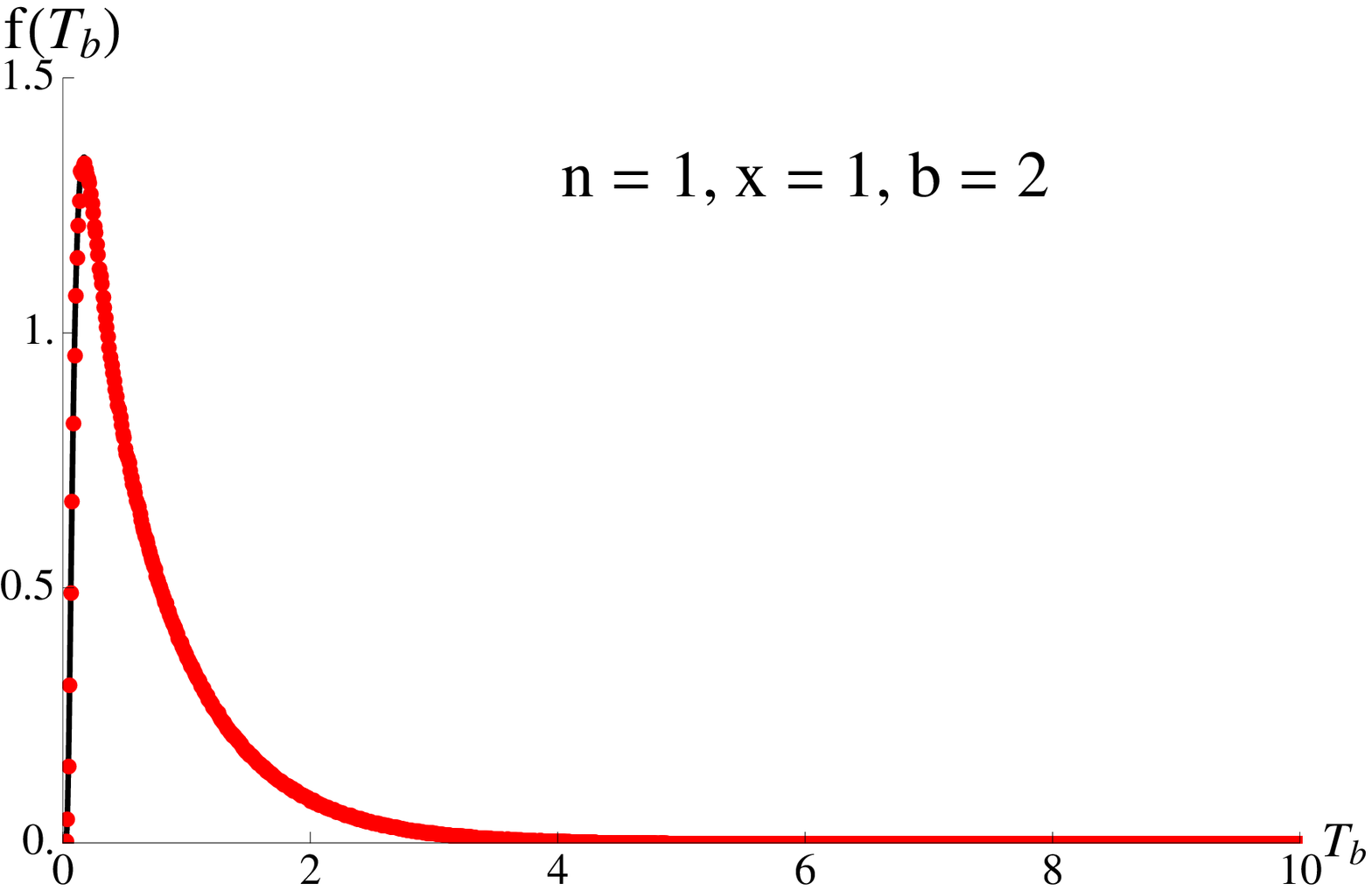}
\caption{\label{fig:fpt_entrance}
First passage time PDF $f(T_b)$ when the origin is an entrance boundary.
The parameter $n$, the starting position $x$, and the upper boundary $b$ are
given in the insets; the diffusion coefficient $D$ is 1. The analytical results
(lines) are perfectly covered by the normalized histograms obtained from
simulation (circles).}
\end{figure}

Fig.~\ref{fig:fpt_entrance} shows the analytical results obtained with Wolfram
Mathematica 7.0 by truncating the sum in Eq.~(\ref{eq:solution}) after the
first 200 terms, and normalized histograms generated with 10 million simulation
runs done as explained in Sec.~\ref{sec:simulation}. The agreement is perfect.
The CPU time needed for an analytical curve is a few seconds, while that for a
histogram with 10 million runs, which is the number used for all histograms in
this paper, ranges from a few minutes to two days, depending on the time step,
the starting position, and the upper boundary $b$. We used the \texttt{ran1}
uniform random number generator \cite{Press2003} and the GNU C++ compiler (g++)
version 4.1.2 with the -O3 optimization option on a 2.2 GHz AMD Athlon 64
``Winchester'' processor with Fedora Core 7 Linux.

\subsubsection{Exit boundary}

For $n \leq -1 $, i.e.\ $\nu \geq 1$, the zero level is an exit boundary and
it is impossible to reach any interior point $b$, provided that the starting
point of the process is sufficiently close to the boundary. This, repeating the
arguments of Sec.~\ref{sec:heurclass}, corresponds to a collapse of the PDF
to a delta function $\delta(x)$ in a finite time. Hence in general the
first-passage time with respect to $x = b > 0$ will diverge.

However, with the absorbing boundary at the origin, where naturally $u(0) = 0$,
and imposing an absorbing boundary condition at the upper boundary, $u(b) = 0$,
we have a boundary value problem with a solution that is the PDF of the
first-exit time $T_{0,b} = \min\{T_0,T_b\}$ from the interval $(0,b)$. 

The first terms in Eqs.~(\ref{eq:generalsolution1}) or
(\ref{eq:generalsolution2}) vanish in the limit $x \to 0$, whereas the second
terms {\it diverge}. Therefore the constant $B$ must be zero in order to
fulfill $u(b) = 0$. As in the case of an entrance boundary, the eigenvalues
are determined by the condition $J_\nu(\sqrt{\alpha_k}b) = 0$ and the
eigenfunctions are given by Eq.~(\ref{eq:eigenfunctions1}).

\begin{figure}[htbp]
\includegraphics[width=0.49\columnwidth]{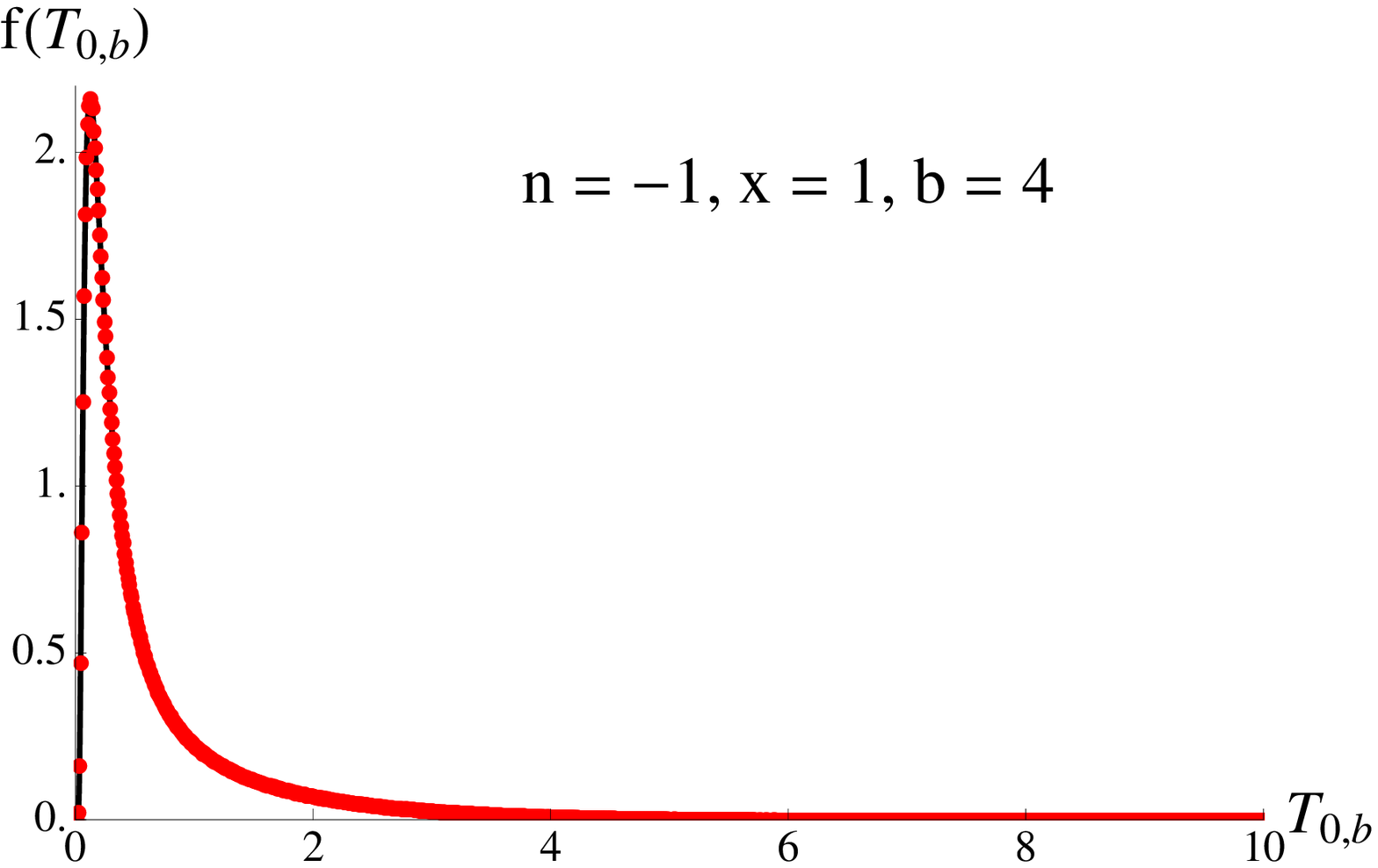}
\includegraphics[width=0.49\columnwidth]{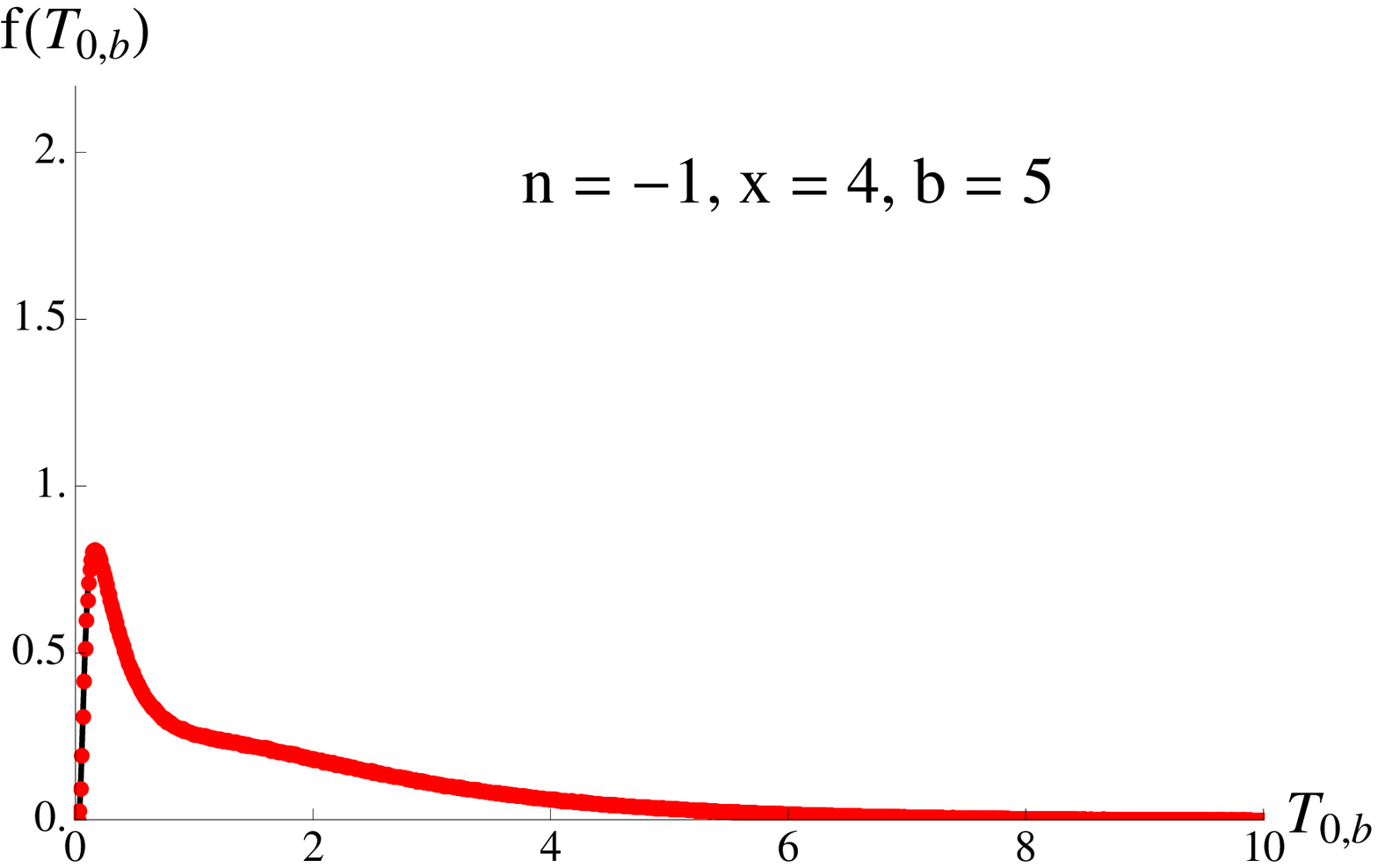}
\includegraphics[width=0.49\columnwidth]{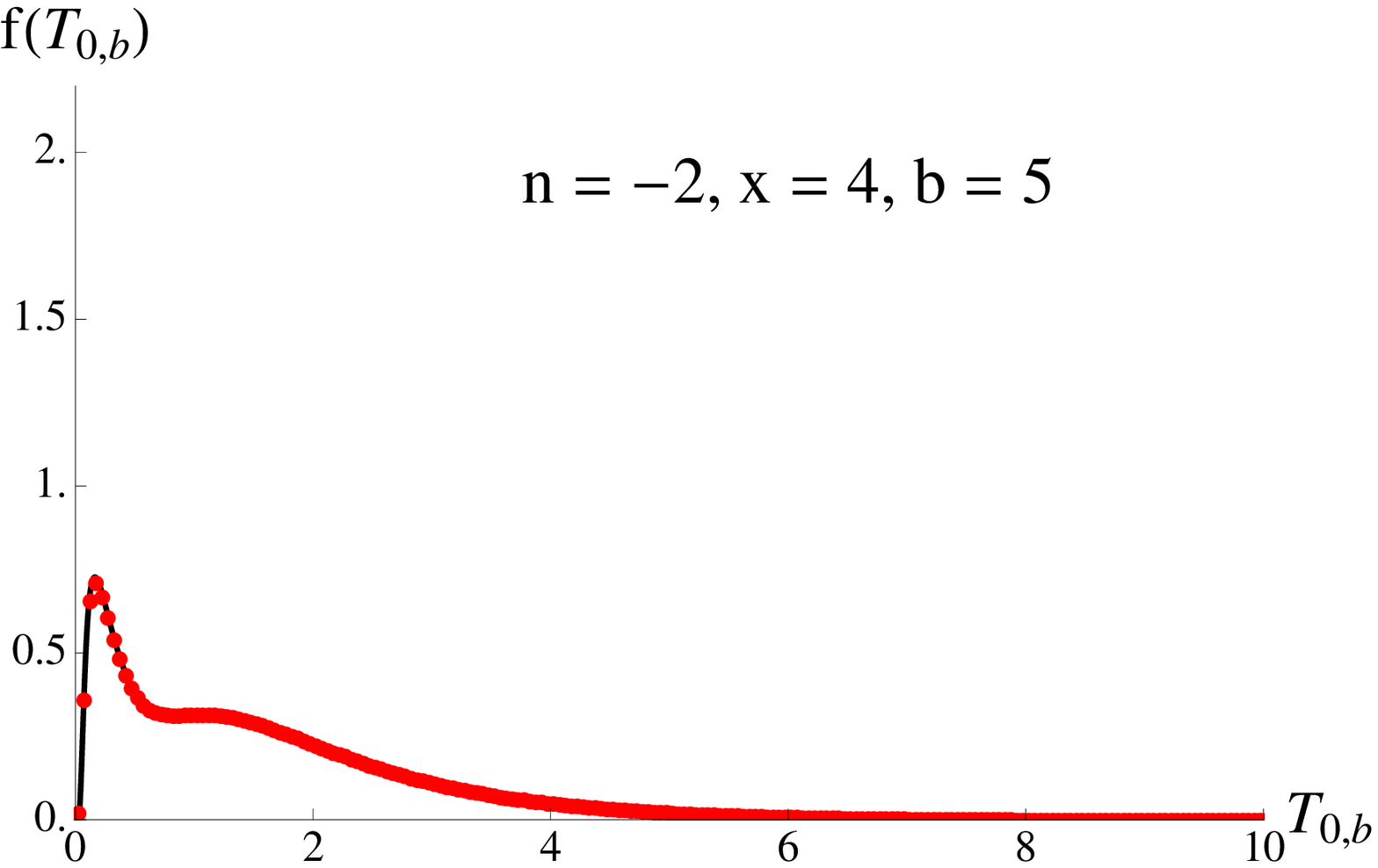}
\includegraphics[width=0.49\columnwidth]{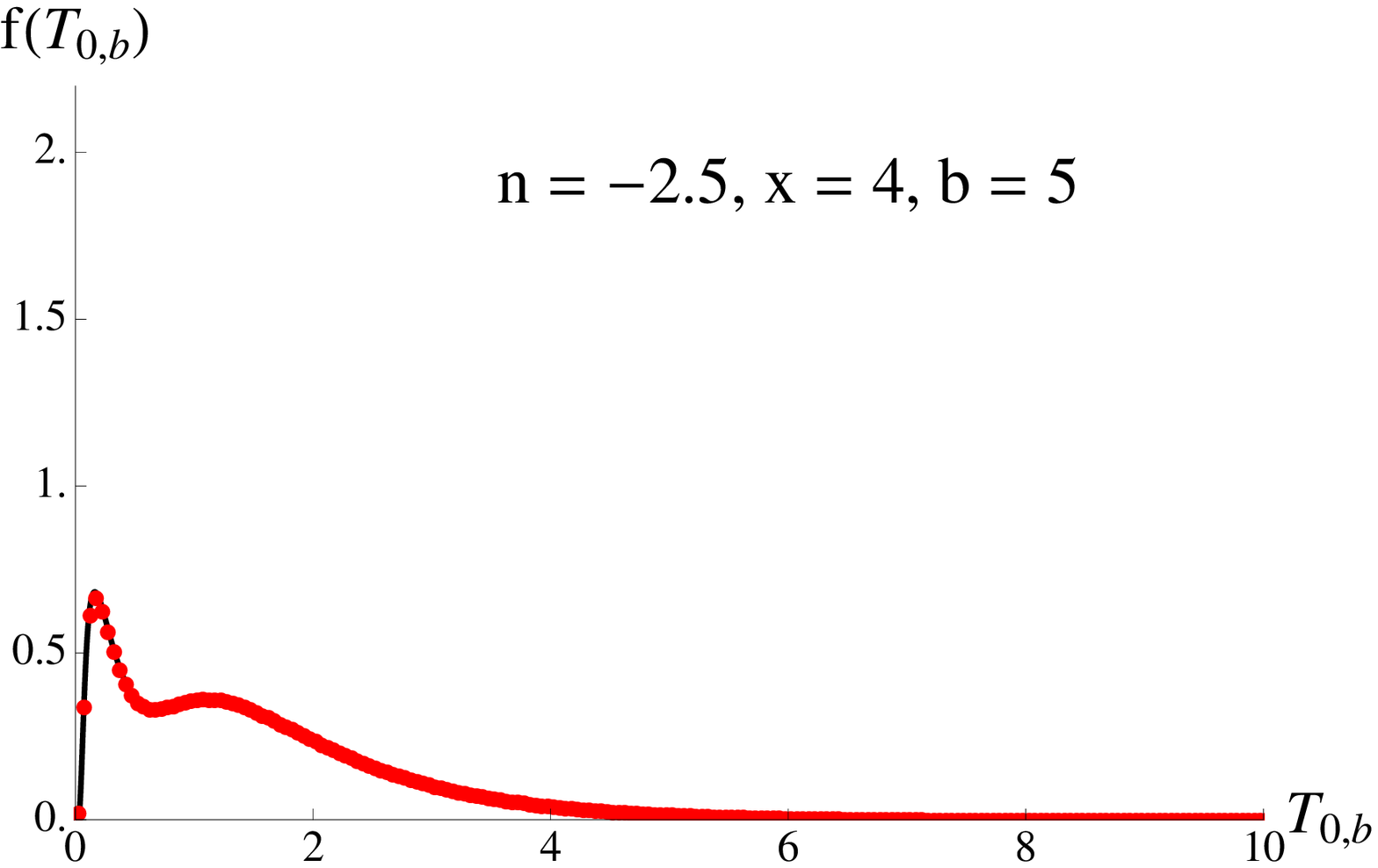}
\caption{\label{fig:fpt_exit}
First exit time PDF $f(T_{0,b})$ when the origin is an exit boundary.
The parameter $n$, the starting position $x$, and the upper boundary $b$ are
given in the insets; the diffusion coefficient $D$ is 1. The analytical results
(lines) are perfectly covered by the normalized histograms obtained from
simulation (circles).}
\end{figure}

In Fig.~\ref{fig:fpt_exit} the theoretical curves are again compared with the
results obtained numerically. It is interesting to notice that the PDF is
bimodal for a range of parameters.

\subsubsection{Regular boundary}

For $-1 < n < 1$, i.e.\ $0 < \nu < 1$, the origin is a regular boundary, and
in accordance with Karlin and Taylor \cite{Karlin1981} it is possible to impose
different boundary conditions in a consistent way.

\begin{figure}[htbp]
\includegraphics[width=0.49\columnwidth]{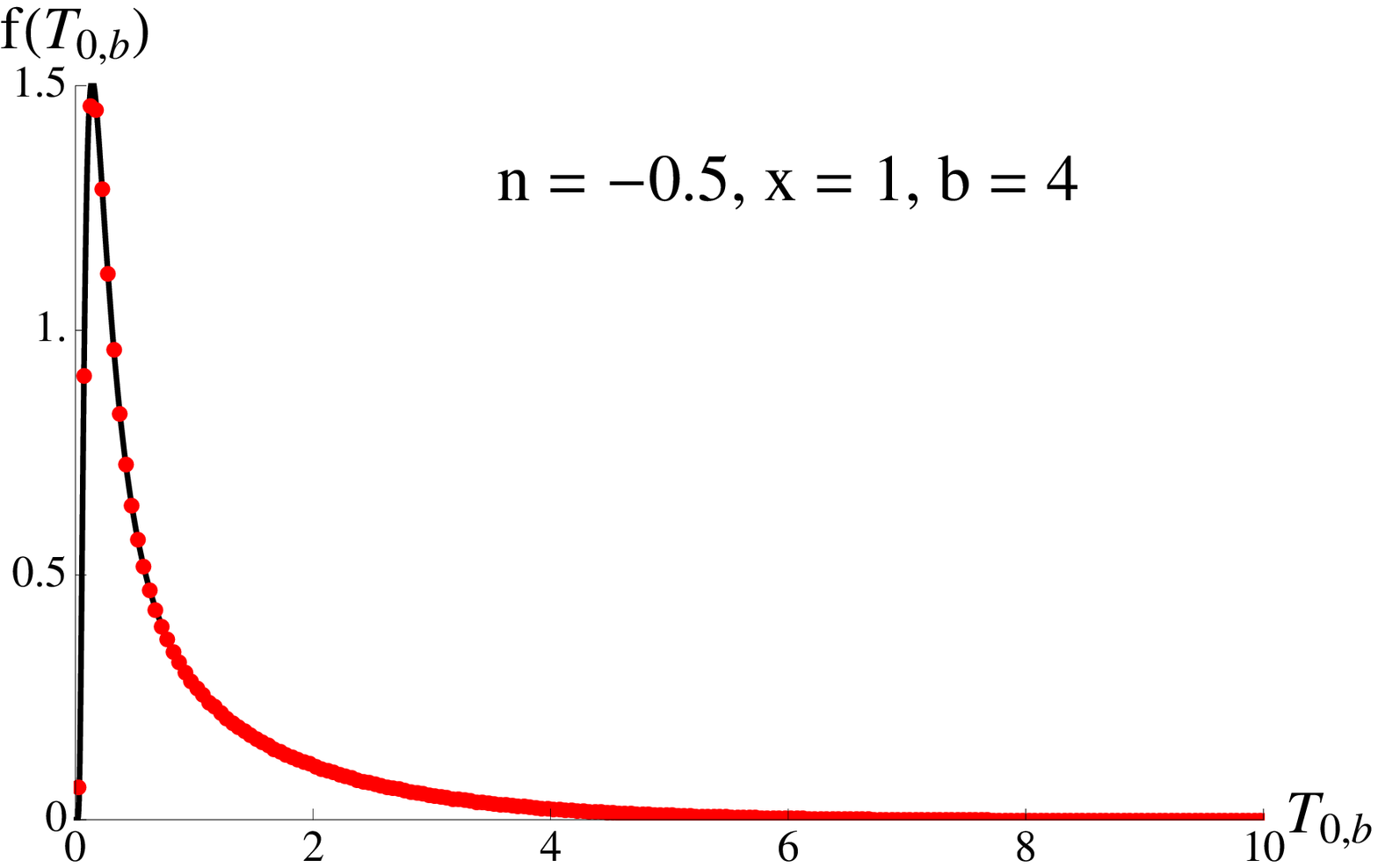}
\includegraphics[width=0.49\columnwidth]{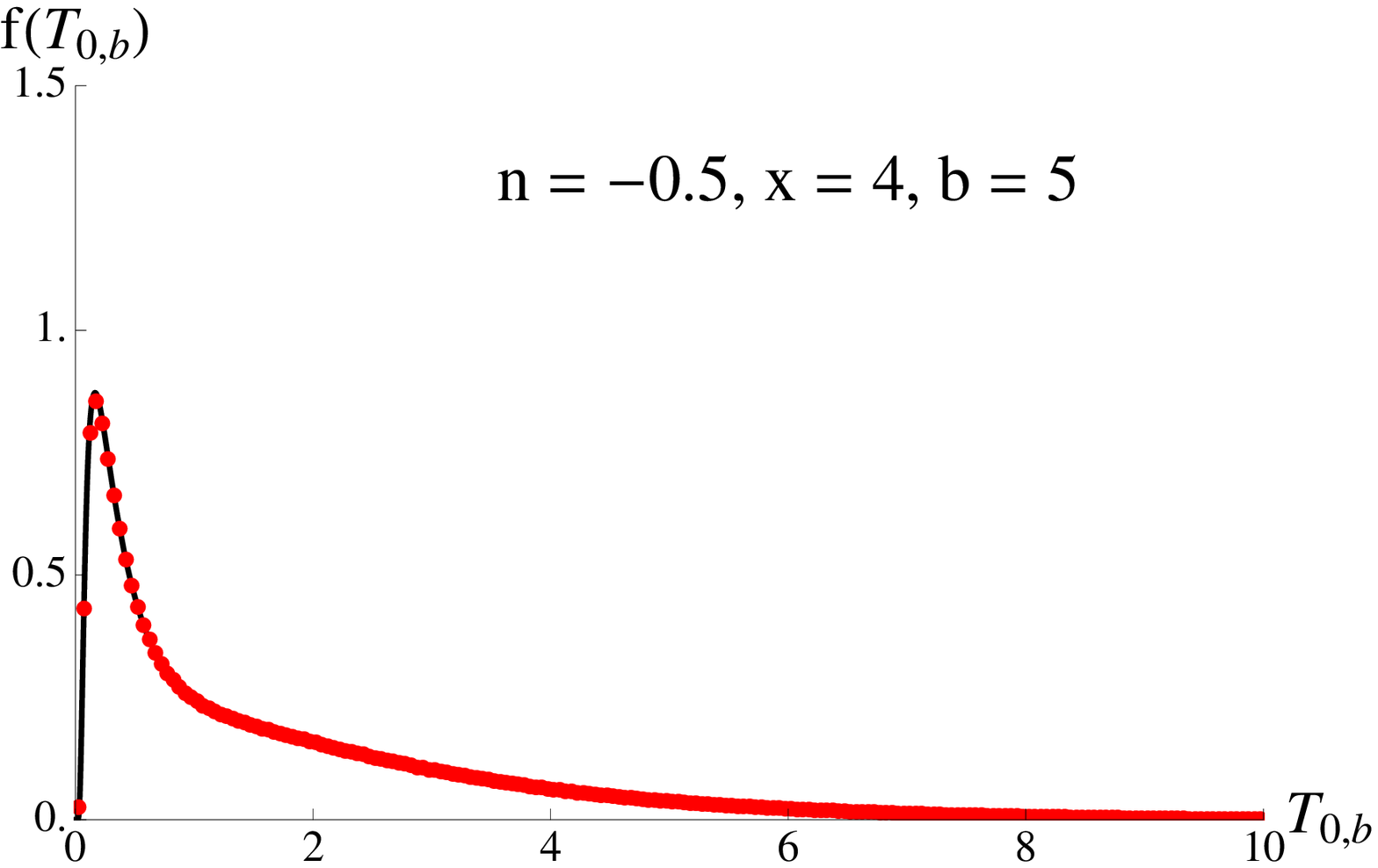}
\includegraphics[width=0.49\columnwidth]{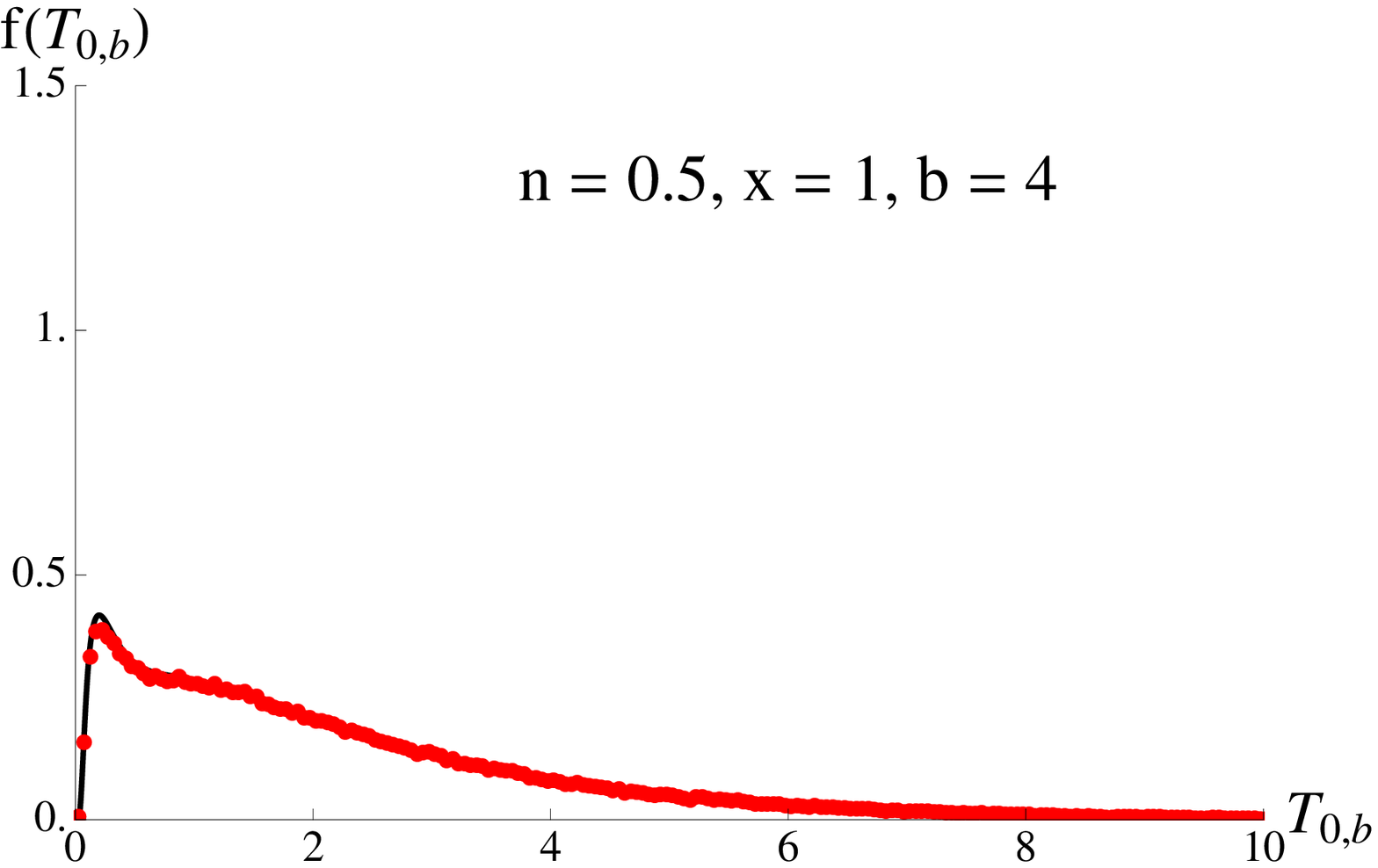}
\includegraphics[width=0.49\columnwidth]{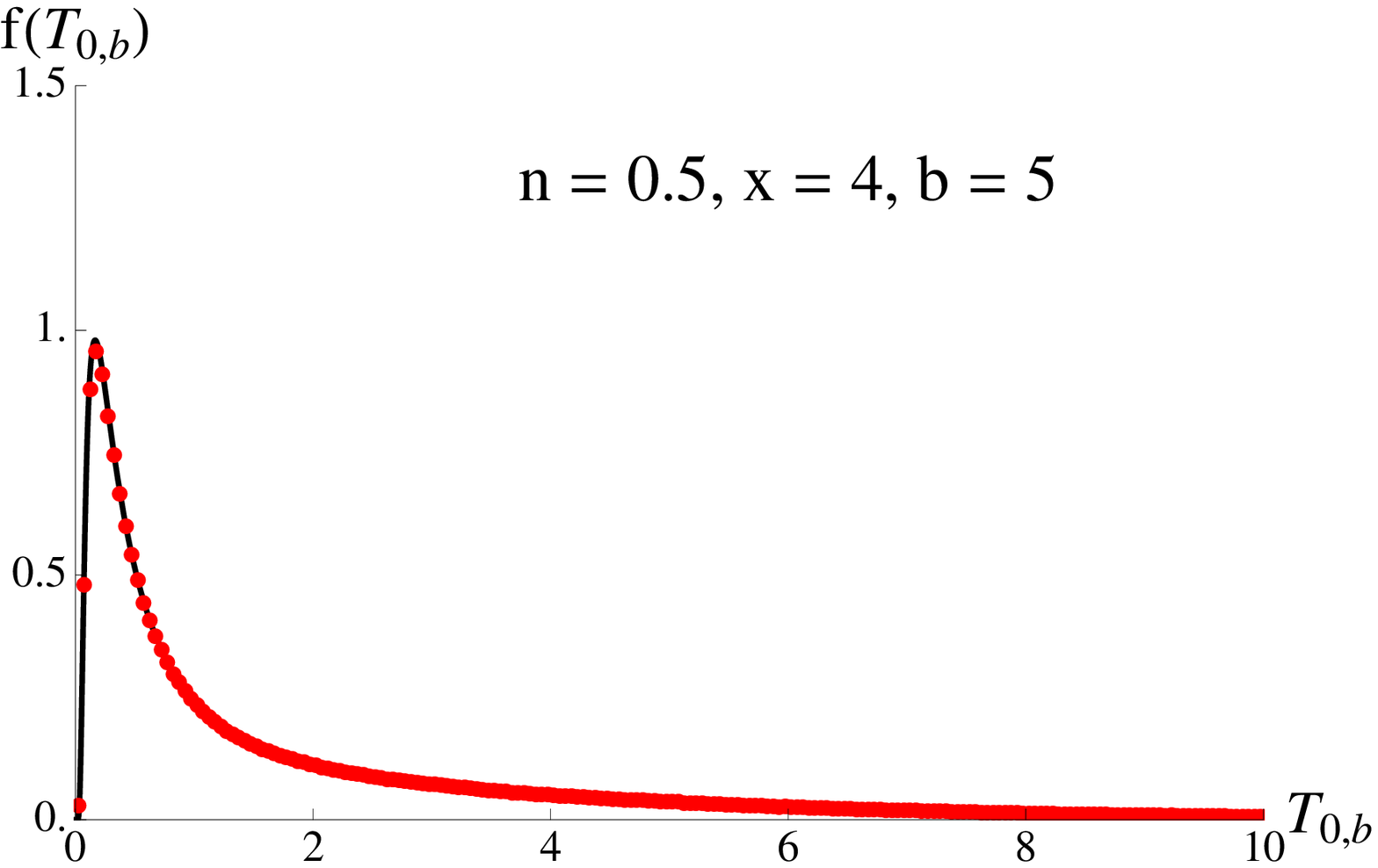}
\caption{\label{fig:fpt_regular_first_exit}
First exit time PDF $f(T_{0,b})$ when the origin is a regular boundary imposed
to be absorbing. The parameter $n$, the starting position $x$, and the upper
boundary $b$ are given in the insets; the diffusion coefficient $D$ is 1.
The analytical results (lines) are perfectly covered by the normalized
histograms obtained from simulation (circles).}
\end{figure}

Imposing an absorbing boundary condition at the origin gives the PDF of the
first-exit time. The eigenvalues are computed in the same way as for an exit
boundary at the origin, and the eigenfunctions are again given by
Eq.~(\ref{eq:eigenfunctions1}), which was also the result for the entrance
boundary in the case of negative and integer $\nu$.
Fig.~\ref{fig:fpt_regular_first_exit} shows that the theoretical results agree
perfectly with the simulations.

\begin{figure}[htbp]
\includegraphics[width=0.49\columnwidth]{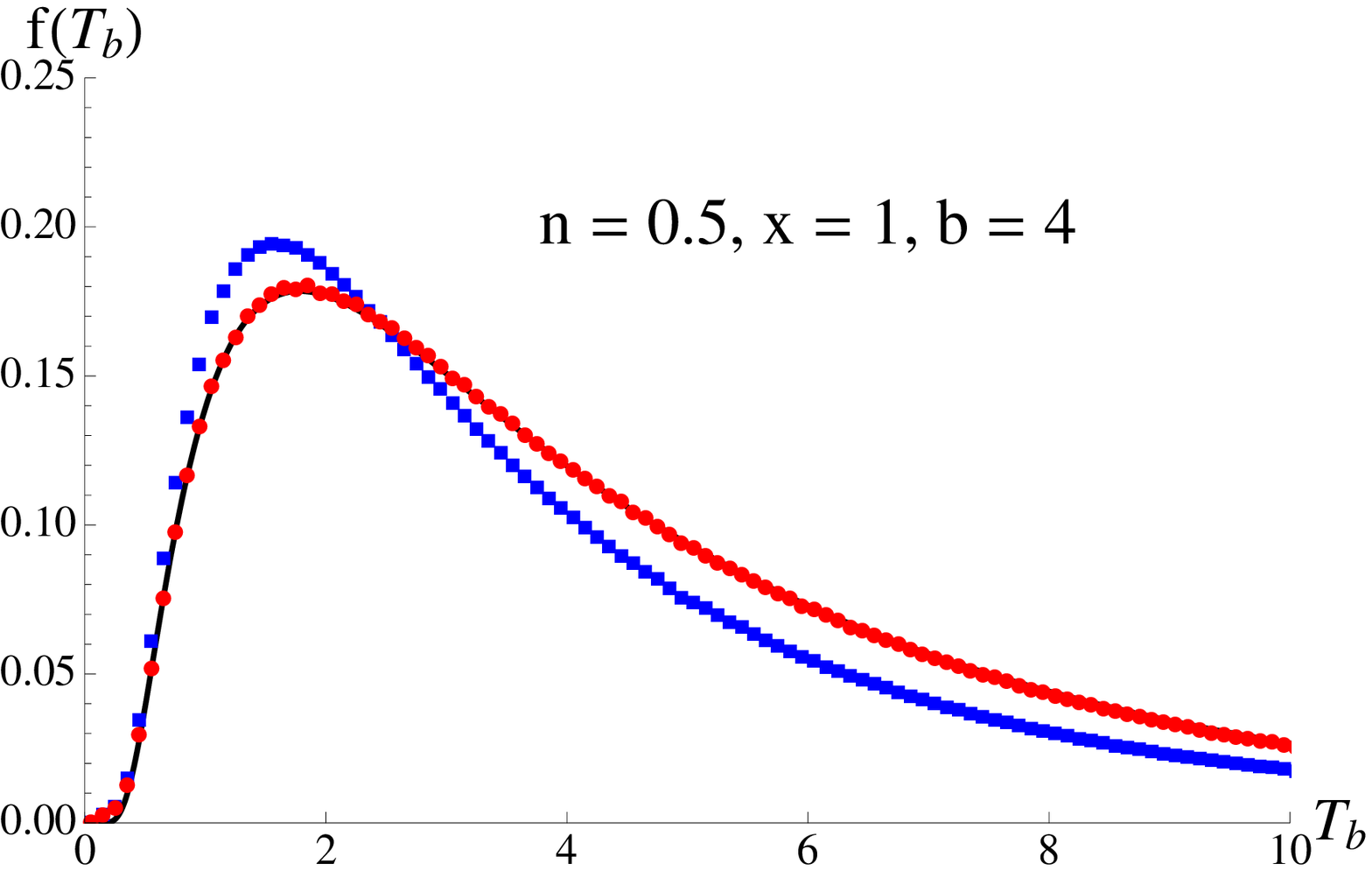}
\includegraphics[width=0.49\columnwidth]{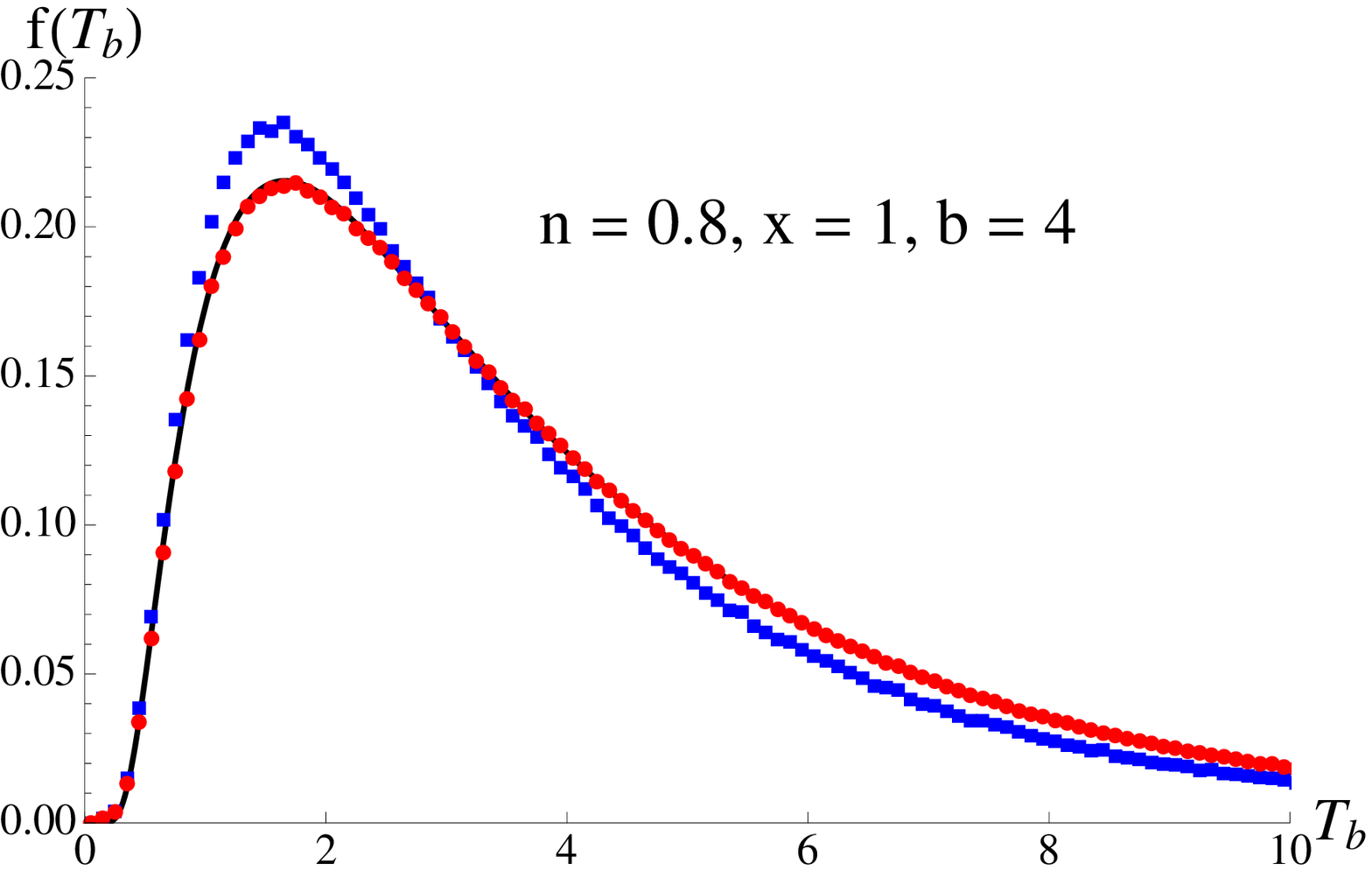}
\caption{\label{fig:fpt_regular}
First passage time PDF $f(T_b)$ when the origin is a regular boundary imposed
to be reflecting. The parameter $n$, the starting position $x$, and the upper
boundary $b$ are given in the insets; the diffusion coefficient $D$ is 1. The
analytical results (lines) are perfectly covered by the normalized histograms
obtained from simulation (circles). Mismatching simulation results (squares)
arise if zero crossings are allowed.}
\end{figure}

It is also possible to impose a reflecting boundary condition at the origin,
and hence the eigenfunctions are computed in the same way as in the case of an
entrance boundary, of course inserting the respective value of $n$. For a few
values of $n$ in the range $0 < n < 1$ we have compared the theoretical curves
resulting from the assumption that the origin is reflecting with histograms
from simulations where we have allowed zero crossings; see the squares in
Fig.~\ref{fig:fpt_regular}. It appears that for small times there is a good
agreement, whereas for larger times there are differences: the maxima of the
histograms are higher than predicted by the theory, and the tails obtained by
simulations are flatter than the theoretical tails. So we can clearly conclude
that the origin is not naturally reflecting for this range of $n$, but, as one
can see in Fig.~\ref{fig:fpt_regular}, total reflection is approached when $n$
is approaching the limit where the origin is an entrance boundary, namely
$n = 1$.

\begin{figure}[htbp]
\includegraphics[width=0.7\columnwidth]{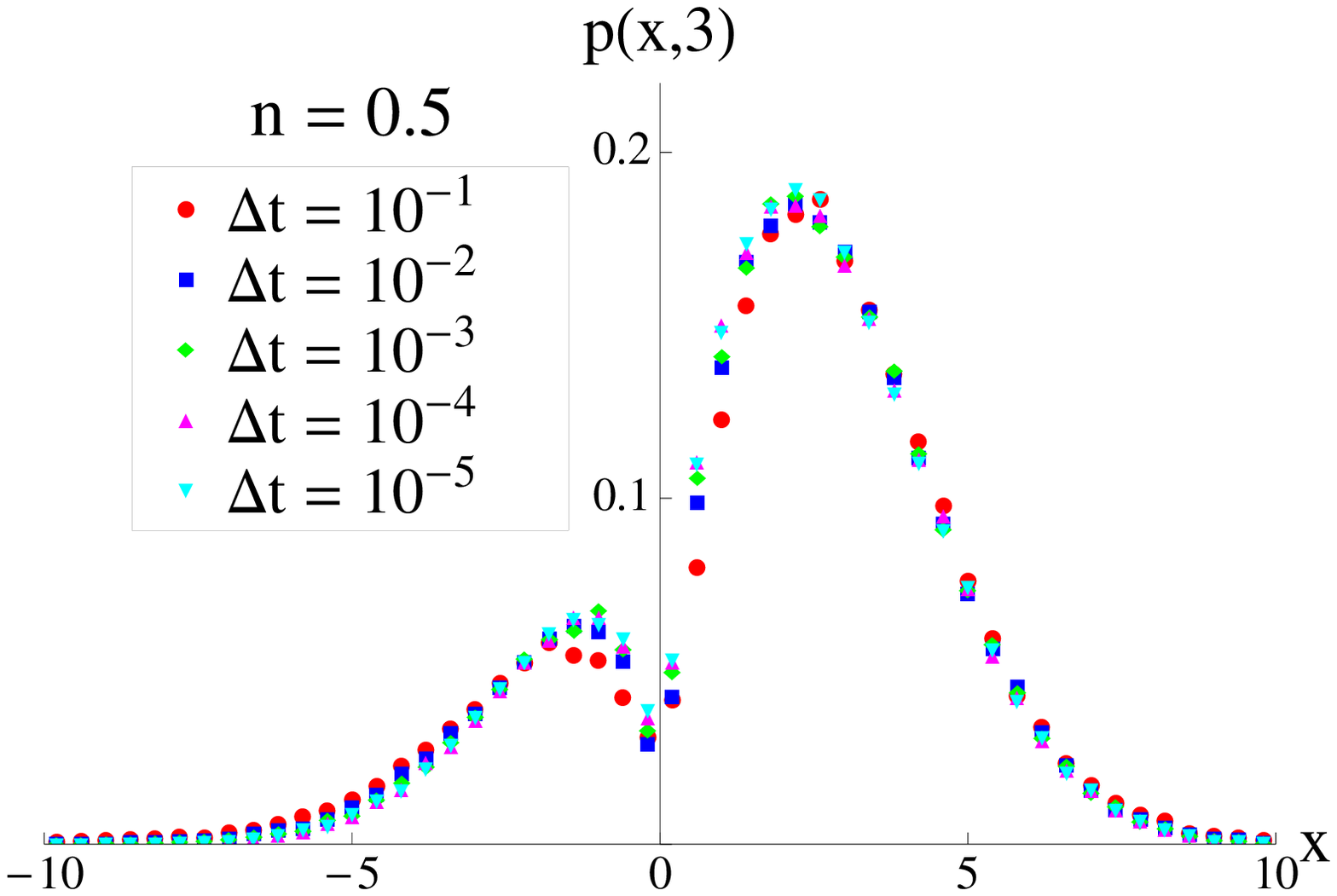}
\includegraphics[width=0.7\columnwidth]{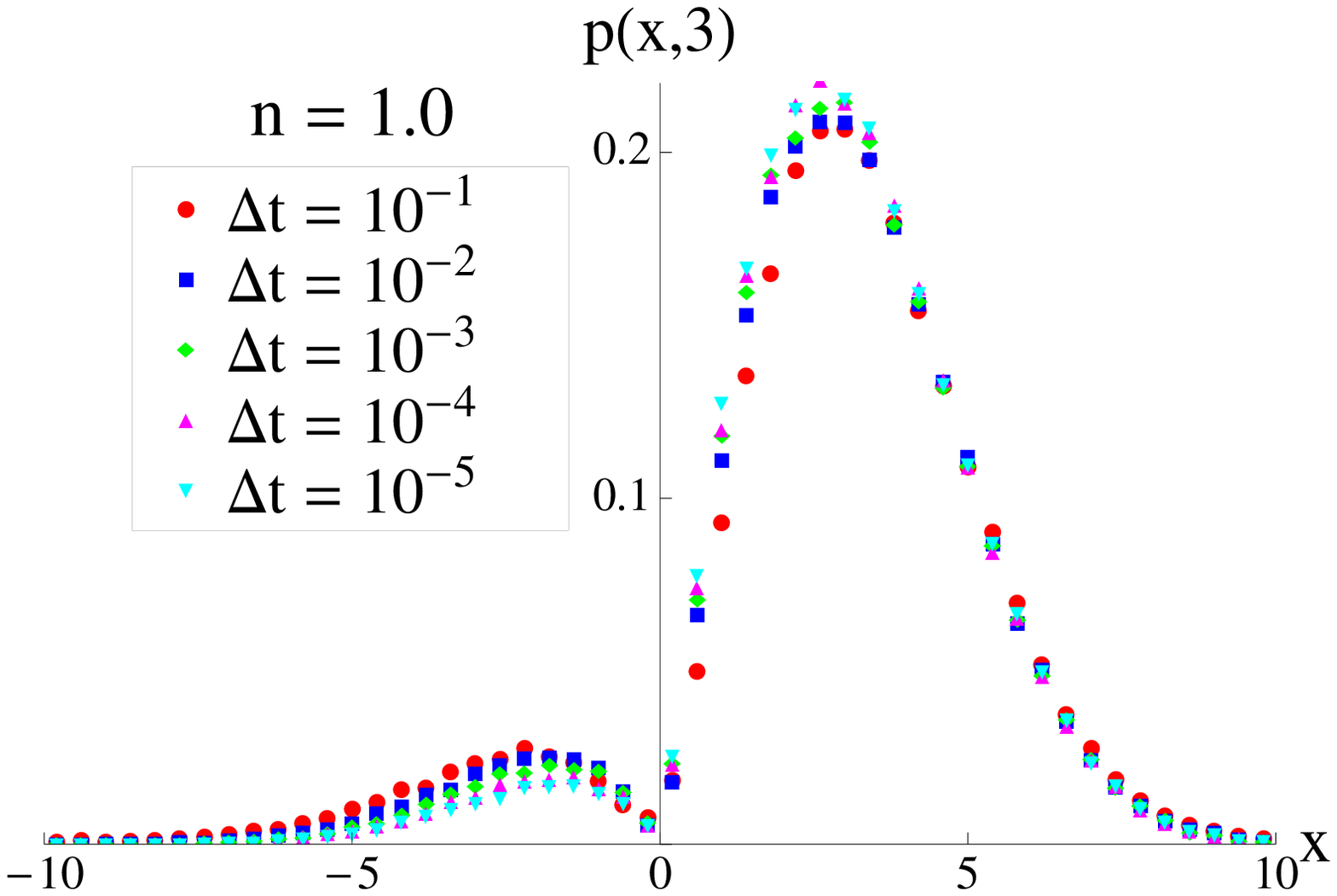}
\caption{\label{fig:fpt_regular_pd}
PDFs of the stochastic process $X(t)$ at $t = 3$ from simulation with different
integration time step $\Delta t$ for $n = 0.5$ (left) and $n = 1.0$ (right);
the starting position $x_0$ and the diffusion coefficient $D$ are 1. For
$n = 0.5$ the peak in the negative domain increases with decreasing time step,
whereas for $n = 1$ it decreases. This suggests that in the latter case, where
the origin is an entrance boundary, the zero crossings are an artifact due to
the discretization of time, whereas in the former case, where the origin is
regular, the zero crossings are genuine.}
\end{figure}

To explain this phenomenon we recall what we have mentioned earlier: one might
think intuitively that zero crossings are not possible for non-zero values
of $n$ since the drift term explodes near the origin, and the latter either
reflects or absorbs the process for all times. However, applying Feller's
formal classification scheme one can see that zero-crossings are actually
allowed for a regular boundary at the origin, i.e.\ $-1 < n < 1$. This is
further confirmed by Fig.~\ref{fig:fpt_regular_pd}.

\begin{figure}[htbp]
\bigskip
\includegraphics[width=0.7\columnwidth]{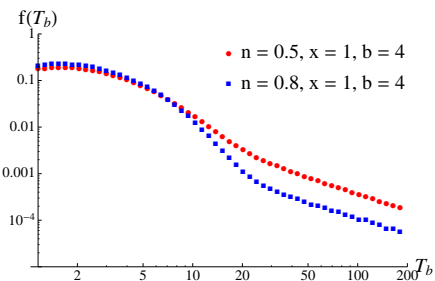}
\caption{\label{fig:fpt_regular_logplot}
Normalized histograms of first-passage times with respect to the level $b$
obtained by simulation when the origin is a regular boundary. The parameter
$n$, the starting position $x$, and the upper boundary $b$ are given in the
inset; the diffusion coefficient $D$ is 1. The tails can be fitted by power
laws with exponents -1.11 and -1.17, respectively.}
\end{figure}

Knowing this and the fact that according to Karlin and Taylor \cite{Karlin1981}
a process can spend a finite time in the vicinity of a regular boundary, we
can explain the plots qualitatively. The paths that are able to escape the
influence of the origin will quickly hit the boundary $b$ following the same
rules as for the entrance boundary; they are basically driven by the drift
term. The deviation in the tail of the PDFs is due to the positive amount of
time spent in the vicinity of the origin, which is called the sticky boundary
phenomenon \cite{Karlin1981}, and to the multiple zero crossings.
Fig.~\ref{fig:fpt_regular_logplot} shows a logarithmic plot of the
first-passage time PDFs obtained by simulations, and one can see that the
latter are heavy-tailed, i.e.\ they exhibit a power-law decay for long times.
This is in contrast to the exponential decay obtained for the other types of
boundary.

However, there is a good agreement between theory and simulation if we impose a
reflecting origin in the simulation too, meaning that we consider the origin as
a hard reflecting wall; see the circles in Fig.~\ref{fig:fpt_regular}.

For $-1 < n < 0$ the first-passage times diverge if we do not impose any
artificial boundary condition, since the drift term is always negative if
$X_t > 0$ and positive if $X_t < 0$, meaning that the process is always
attracted, but not totally absorbed, by the origin. On the other hand,
imposing total absorption at the origin corresponds to the computation of the
PDF of the first-exit times as shown in Fig.~\ref{fig:fpt_regular_first_exit}. 

It is interesting to note that the case $n = 0$, i.e.\ the Wiener process,
belongs to this class. The origin is not a singular point and the
first-passage time PDF with respect to a level $b$ starting at $x_0$ is
given by Eq.~(\ref{eq:WienerFPTdensity}), which for long times is a power law
with exponent $-3/2$. 

\section{Conclusions}

We have computed first-passage and first-exit time PDFs for a stochastic
process with applications in many physical, chemical, biological, economical
and financial problems. Depending on the nature of the boundary at the origin,
we have found analytical solutions for the first-passage and first-exit time
PDFs for all cases, except for the first-passage time PDF in the case of a
regular boundary at the origin. In the latter case we have found an analytical
solution for the first-exit time PDF and approximations for the first-passage
time PDF for short times. For this specific stochastic process regularity of
the boundary at zero can include behaviours ranging from total absorption to
total reflection, with intermediate behaviours like elastic and sticky
boundaries \cite{Karlin1981}. It is interesting that sticky boundaries may be
applied e.g.\ to simulate the partial adsorption of polymer molecules to walls
and for the modeling of solvent quality \cite{Peters2002}. In possible future
projects this could be investigated more thoroughly and regarded from the
perspective of interactions between molecules and boundary surfaces, which
is closely connected to another project of two of us \cite{Stillings2008,
Martin2010}, where discotic liquid crystals confined in cylindric geometries
\cite{Caprion2009} are studied via molecular dynamics simulations.

\section*{Acknowledgements}

We thank R.~Mannella, D.~Marazzina, B.~Schmitt, and A.~Zettl for useful
discussions and observations.

\section*{Appendix}

We prove the Fourier-Bessel expansion given in Eq.~(\ref{eq:Fourier-Bessel2}).
If a function $f(z)$ is represented in an orthogonal basis of Bessel functions
of the first kind $J_\nu(j_kz)$, where $j_k$ is the $k$th zero of $J_\nu(z)$,
i.e.\ $J_\nu(j_k) = 0$,
\begin{equation}
f(z) = \sum_{k=1}^\infty c_k J_\nu(j_kz),
\end{equation}
and the orthogonality relation is given by Eq.~(\ref{eq:Bessel_orthogonality}),
the $l$th coefficient $c_l$ can be obtained from the scalar product of $f(z)$
with the $l$th basis set element $J_\nu(j_lz)$,
\begin{eqnarray}
\int_0^1 J_\nu(j_lz) f(z) z \, dz
& = & \sum_{k=1}^\infty c_k \int_0^1 J_\nu(j_kz) J_\nu(j_lz) z\, dz \nonumber\\
& = & \sum_{k=1}^\infty \frac{c_k}{2} J_{\nu+1}^2(j_k) \delta_{lk} \nonumber \\
& = & \frac{c_l}{2} J_{\nu+1}^2(j_l),
\end{eqnarray}
resulting in
\begin{equation}
c_k = \frac{2}{J_{\nu+1}^2(j_k)} \int_0^1 J_\nu(j_kz) f(z) z \, dz
= \frac{2I_k}{J_{\nu+1}^2(j_k)}.
\end{equation}
For $f(z) = z^{-\nu}$
\begin{equation}
I_k = \int_0^1 J_\nu(j_kz) z^{1-\nu} \, dz.
\end{equation}
In order to exploit the equation \cite{Wolfram2010}
\begin{equation}
\int J_\nu(z) z^{1-\nu} \, dz = - J_{\nu-1}(z) z^{1-\nu},
\end{equation}
we substitute $j_kz = \alpha$ and get
\begin{eqnarray}
I_k
& = & j_k^{\nu-2} \int_0^{j_k} J_\nu(\alpha)\alpha^{1-\nu}\,d\alpha \nonumber\\
& = & j_k^{\nu-2} \left[ - J_{\nu-1}(\alpha)\alpha^{1-\nu} \right]_0^{j_k}
      \nonumber \\
& = & j_k^{\nu-2} \left[ \lim_{\alpha \to 0} J_{\nu-1}(\alpha) \alpha^{1-\nu}
      - J_{\nu-1}(j_k) j_k^{1-\nu} \right].
\end{eqnarray}
The limit is
\begin{eqnarray}
\lim_{\alpha \to 0} \alpha^{1-\nu} J_{\nu-1}(\alpha)
& = & \lim_{\alpha \to 0} \alpha^{1-\nu} \sum_{l=0}^\infty \frac{(-1)^l}
      {l!\,\Gamma(l+\nu)} \left( \frac{\alpha}{2} \right)^{2l+\nu-1}\nonumber\\
& = & \sum_{l=0}^\infty \frac{(-1)^l\,2^{1-\nu-2l}}{l!\,\Gamma(l+\nu)}
      \lim_{\alpha \to 0} \alpha^{2l} \nonumber \\
& = & \frac{2^{1-\nu}}{\Gamma(\nu)},
\end{eqnarray}
yielding
\begin{equation}
I_k = j_k^{\nu-2} \left[ \frac{2^{1-\nu}}{\Gamma(\nu)}
- \frac{J_{\nu-1}(j_k)}{j_k^{\nu-1}} \right].
\end{equation}
Thus
\begin{equation}
z^{-\nu} = \sum_{k=1}^{\infty} \left[ \frac{(2/j_k)^{2-\nu}}{\Gamma(\nu)}
- \frac{2 J_{\nu-1}(j_k)}{j_k}\right] \frac{J_\nu(j_kz)}{J_{\nu+1}^2(j_k)}.
\end{equation}
Subtracting the Fourier-Bessel expansion of $z^\nu$,
Eq.~(\ref{eq:Fourier-Bessel1}), the second term in square braces cancels
out because of the recurrence identity $J_{\nu-1}(j_k) + J_{\nu+1}(j_k)
= 2\nu J_\nu(j_k)/j_k = 0$ that we have already used to simplify
Eq.~(\ref{eq:scalarproducts1}) to Eq.~(\ref{eq:scalarproducts1integer}).

\bibliography{paper}

\end{document}